# Water Stress on U.S. Power Production at Decadal Time Horizons

**Project Title: "Future U.S. Water Availability & Quality Study"**

**Principal Investigator and Point of Contact:**
**Auroop R. Ganguly**
Address: Sustainability and Data Sciences Laboratory (SDS Lab),
Civil and Environmental Engineering, Northeastern University,
360 Huntington Avenue, 400SN, Boston, MA-02115
Phone: 617-373-6005; Email: a.ganguly@neu.edu

**Prepared By:**
Poulomi Ganguli, Devashish Kumar, and Auroop R. Ganguly
SDS Lab, Northeastern University, Boston, MA-02115

**Prepared For:**
Advanced Research Projects Agency – Energy
United States Department of Energy
1000 Independence Avenue SW
Washington, DC 20585

Northeastern University

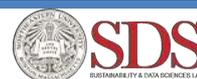

# TABLE OF CONTENTS

Northeastern University                    SDS

# ACKNOWLEDGEMENT

Funding was primarily provided by the United States Department of Energy's (U.S. DOE) Advanced Research Projects Agency – Energy (ARPA-E) through the DOE Purchase Order #DE-AR0000482 entitled "Future U.S. Water Availability & Quality Study", and complemented by Northeastern University's Office of the Provost and College of Engineering through a Tier II grant. ARPA-E helped with problem definition, power plant related data procurement, and evaluation of solution strategies. Climate data were obtained from the Program for Climate Model Diagnosis and Intercomparison (PCMDI) archive of the U.S. Department of Energy's Lawrence Livermore National Laboratory (LLNL). Hydrologic data were obtained from the website of United States Geological Survey (USGS). Population data were taken from the United States Census Bureau. Northeastern University's College of Engineering and Department of Civil and Environmental Engineering provided software such as ArcGIS and the required computational resources. Janet Yun, an undergraduate student in Civil and Environmental Engineering at Northeastern University, helped during the initial analysis phase through an independent study program. We are grateful to Tyler Hall, a Northeastern University Scholar and undergraduate student in the Department of Mechanical Engineering, for his technical and editorial help with this report write-up. The work was performed at the Sustainability and Data Sciences Laboratory (SDS Lab) of the Department of Civil and Environmental Engineering at Northeastern University.

# Executive Summary

Thermoelectric power production at risk, owing to current and projected water scarcity and rising stream temperatures, is assessed for the contiguous United States at decadal scales. Regional water scarcity is driven by climate variability and change, as well as by multi-sector water demand. While a planning horizon of zero to about thirty years is occasionally prescribed by stakeholders, the challenges to risk assessment at these scales include the difficulty in delineating decadal climate trends from intrinsic natural or multiple model variability. Current generation global climate or earth system models are not credible at the spatial resolutions of power plants, especially for surface water quantity and stream temperatures, which further exacerbates the assessment challenge. Population changes, which are difficult to project, cannot serve as adequate proxies for changes in the water demand across sectors. The hypothesis that robust assessments of power production at risk are possible, despite the uncertainties, has been examined as a proof of concept. An approach is presented for delineating water scarcity and temperature from climate models, observations and population storylines, as well as for assessing power production at risk by examining geospatial correlations of power plant locations within regions where the usable water supply for energy production happens to be scarcer and warmer. Our analyses showed that in the near term, more than 200 counties are likely to be exposed to water scarcity in the next three decades. Further, we noticed that stream gauges in more than five counties in the 2030s and ten counties in the 2040s showed a significant increase in water temperature, which exceeded the power plant effluent temperature threshold set by the EPA. Power plants in South Carolina, Louisiana, and Texas are likely to be vulnerable owing to climate driven water stresses. In all, our analysis suggests that under various combinations of plausible climate change and population growth scenarios, anywhere between 4.5 and 9 quads of delivered electricity (from existing plants) would be generated in counties that are at risk of water scarcity and/or unacceptably high stream temperatures.

**Process Flow:** Water stress, owing to lower availability and rising temperatures, directly puts thermoelectric power production at risk. Climate change and variability impacts the hydrologic cycle and hence water supply, as well as stream temperature. Changes in population and multi-sector water usage impact water demand. While reduction in water

Northeastern University                                           SDS

supply and increase in demand lead to water availability stresses on power plant operations, higher water temperature leads to water quality stresses. Power plant attributes, including type, capacity and location are required for exposure analysis along with information about regional water resources to understand overall vulnerability. Water stress from scarcity and warming needs to be correlated with power plant exposure and resilience to develop risk assessments. The challenges stem from the scales and horizons of projections as well as the complexity of the risk and decision tradeoff space. The concept is shown in Figure ES1.

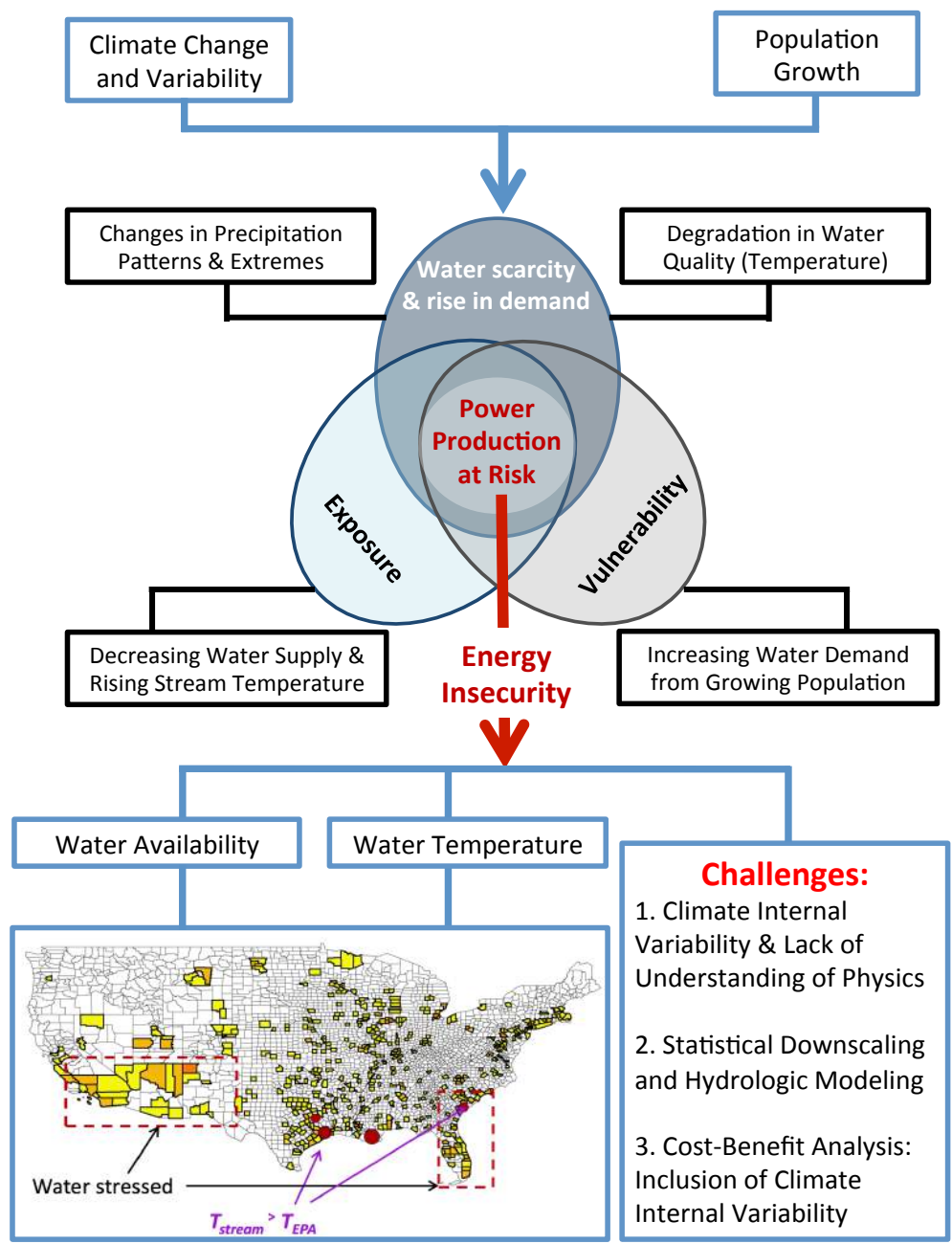

**Figure ES1** Process Flow Diagram

# Chapter 1: Introduction

Energy security, water sustainability, and climate change are among the most urgent and imminent threats facing the nation and our global society. In the context of thermoelectric power production, the three areas interact intimately, leading to what may be viewed as a complex 'energy-water-climate nexus'. The impact of climate change on freshwater availability (and hence on energy production) in the future has been studied before; however, most of these studies focused on mid- to end- of the century time periods and continental-to-global spatial scales. However, a wide spectrum of stakeholders across private and public sectors have been stressing the need to study the multi-sector impacts of climate change over the next 30 years and at local to regional scales. The mismatch between the required temporal and spatial scales for power plant managers and energy planners versus what can be credibly generated from climate or water models and observations leads to major challenges. The cost-benefit tradeoffs related to power plant exposures and vulnerabilities, which need to consider multiple scales of behavior and resource availability, exacerbates the challenge. This report presents a proof-of-concept for addressing these challenges.

## 1.1 Motivation

In the United States in 2011, 91% of the total electricity was produced by nuclear and fossil-fuelled thermoelectric power plants (approximately 9% of electricity was produced by hydroelectric power plants and other renewable sources), which accounted for 40% of all surface water withdrawals in the U.S. for various operations including cooling (U.S. DOE/EPSA-0002, 2014; EIA, 2011). The U.S. electricity demand will grow by 29% (0.9% per year) from 3,826 billion kWh in 2012 to 4,954 billion kWh in 2040 (EIA, 2014). Electricity consumption tends to increase in line with population growth, approximately 0.8% per year from 2012-2040 (EIA, 2014).

Water plays critical roles in the operations of a power plant. Water is withdrawn at various stages such as cooling and condensing the high-pressure steam that drives turbines, for producing energy for the electricity-generating devices, and for constructing power stations (Fthenakis and Kim, 2010). Water used for cooling is the largest fraction of total water use in

Northeastern University SDS

power plants; it represents more than 90% of total water use (U.S. DOE/EPSA-0002, 2014; EIA, 2013).

Power production in the U.S. has already been impacted by diminished water availability and increased stream temperature. For example, in mid-August 2013, one of the reactors in Millstone nuclear power plant in Waterford, Connecticut had to shut down because the temperature of withdrawn water from the Long Island Sound exceeded the reactor's safety limit by about 1°C (*i.e.*, 23.89°C)[1]. During summer 2012, when heat waves spread across Midwest, one of the generators at Powerton coal plant in central Illinois had to temporarily shut down since available water was too warm to effectively cool the hot steam[2].

Change in precipitation patterns due to climate change may increase the vulnerability of existing power plants and can threaten the viability of new energy projects. In addition, stream temperatures are projected to rise over most parts of the United States because of an increase in ambient air temperature due to global warming. An increase in the temperature of intake water reduces the cooling efficiency (Linnerud et al., 2011) and as stream temperature increases and water supply becomes scarce, power plants generate less power than installed capacity, especially during summer. A report (Rogers et al., 2013) from the Union of Concerned Scientists highlighted that by 2040; low flows and high stream temperatures could reduce the power generation capability of once-through cooling system in the New Madrid Coal plants and Mississippi River basin in Southern Missouri. Several regions in U.S. are under extreme droughts or have experienced severe droughts in the recent past; future demands of water in these regions might exceed available supply even in normal years, especially on the West coast. A deeper, quantitative understanding of the potential impacts of all of these consequences from climate change was thus the focus of this work.

## 1.2 Problem Statement

Here we assess how much thermoelectric power production is at risk for the mainland United States (excluding Alaska and Hawaii) at county levels due to projected water scarcity and

---

[1] http://green.blogs.nytimes.com/2012/08/13/heat-shuts-down-a-coastal-reactor/

[2] http://spectrum.ieee.org/energywise/energy/fossil-fuels/collision-between-water-and-energy-is-underway-and-worsening

**Northeastern University**            **SDS**

rising stream temperature. The motivation for this work came from concerns that in certain regions of the U.S., the decreasing availability of freshwater and rising stream temperature has impacted energy production. To meet the growing demands for fresh water in water-stressed regions, such as California, Texas, and Florida, localities are considering using alternative sources of water such as brackish groundwater, seawater, or other sources that require alternative treatment and use more energy. In areas where water temperatures have risen, power plants have had to curtail, and in some cases, completely halt energy production. There is concern that factors such as climate change and population growth could exacerbate these issues, making them more common and widespread. In this work, we perform a first order assessment of the vulnerability of electricity generation under climate change over the next 30 years (2010-2040) with 5-year climatological mean for multiple time windows within the 30-year time periods.

## 1.3 State of the literature, Best Practices, and Key Gaps

The state-of-the art literature and best practices on assessing power production at risk in the U.S. in water scarcity and rising stream temperature are summarized in Tables A1 and A2. Most of the existing studies (Table A1) for power plant risk assessment focused on water withdrawal assuming increasing demand from population and thermoelectric water usage. Future water demands from other sectors were assumed constant. Some of the studies ran hydrological model (such as the Variable Infiltration Capacity [VIC] model) to map water scarcity (van Vliet et al., 2012) in relation to power plant risk. These studies employed climate data from earlier generations of global circulation models (GCMs) and socioeconomic scenarios for future greenhouse gas (GHG) emissions (Blanc et al., 2014; Roy et al., 2012, 2005). The latest generation of GCMs has a better representation of atmospheric chemistry, improved physics, finer spatial resolutions and has shown slight improvements in simulating precipitation patterns. Most of these studies were focused on water availability from multiple climate models and GHG emissions scenarios and did not address the associated uncertainties resulting from various sources. There are three major sources of uncertainty in projected climate variables: model uncertainty (due to the lack of understanding of physics and numerical modeling of atmospheric processes through different parameterization schemes), scenario uncertainty (insufficient knowledge about the amount of GHG that will be emitted in the future), and climate internal variability. Internal variability

Northeastern University

SDS

arises due to the nonlinear chaotic interactions between different components of the earth system; that means that slight differences in the initial atmospheric conditions lead to different future climates. Most previous studies considered model and scenario uncertainties (*e.g.*, Blanc et al., 2014) on the future availability of fresh water but did not consider internal variability and its impact on availability of fresh water for power generation. The importance of role of internal variability has been highlighted (Deser et al., 2014, 2012) especially for the near-term climatology. At decadal time scales internal variability is dominant enough to obscure the trends in model and scenario uncertainties.

Stream temperatures are estimated using stochastic models or determined using deterministic models. State-of-the-art best practices in estimating stream temperature are listed in Table A2. Deterministic models are better suited for scenario analyses, because their numerical formulations are based on the underlying physics of the systems that influence the heat status of the in-stream flows. The main limitations of the deterministic approaches are the substantial requirements of input data and computing resources. Accuracy of the deterministic models is further constrained by the cascading of uncertainties across model parameters. On the other hand, regression models are based on finding statistical relationships between a set of predictors (meteorological and climate variables) and predictands (stream temperature); however, these models may not necessarily guarantee physical consistencies.

## 1.4 Solution Framework

In this study, we compute the projected changes in water availability using precipitation and evapotranspiration data from the latest generation of climate models, the Coupled Model Intercomparison Project phase 5 (CMIP5) (Stocker et al. 2013: IPCC, AR5 Working Group I report). A first order approximation of water availability is computed as the difference between precipitation and evapotranspiration. We have estimated the future changes in water availability from three climate models and two greenhouse gas emissions scenarios. Details about the climate data are discussed in Chapter 2.

On the demand side, multiple sectors compete for the available water. The U.S. Geological Survey (USGS) collects data on water use. In the last survey in 2005 (Kenny et al., 2009), they reported water withdrawal from seven sectors: municipal public and domestic water

Northeastern University

SDS

supply, industrial, mining, livestock, aquaculture, irrigation for agriculture, and thermoelectric cooling for power generation. Fresh water withdrawals for the cooling of thermoelectric power plants constitute the highest proportion (40%) of all sectors, followed by irrigation (36%) and municipal public and domestic water supply (14%). The three sectors together constitute 90% of total water withdrawals while the other four sectors withdraw 10%. Projecting future water use by sectors is challenging and requires many assumptions. Previous work on projecting water withdrawal and supply for future decades in U.S. assumed that water demand will increase only for municipal domestic and public supply and for thermoelectric cooling (together they constitute 54% of withdrawals currently). From 1970-2005, water withdrawal for irrigation and agriculture has remained within a narrow margin or has declined marginally. The other four sectors, viz., industrial, mining, livestock, and aquaculture, presently use only 10% of water; hence, strong assumptions about no-to-small changes in future water demand by these sectors are not likely to change the insights drastically. The projection for municipal public and domestic water supply sector can be directly tied to the increase in population, and total water demand can be computed by multiplying the projected population with per capita consumption. Hence, we can assume that the remaining water (after deduction from public consumption) will be available for power production. Future technological innovations that may reduce the water requirement for energy production are not considered in the present analysis. Further, we have not considered local or regional scale demographic shifts in projected population (*e.g.*, regional migration).

Power production is also affected by water temperature because hot water reduces the efficiency of cooling systems and hence, the amount of energy produced. Further, high water temperatures discharged from power plants also pose threats to aquatic life and ecosystems. Brayton Point Power Station, one of the largest fossil-fuelled power plants in Northeast U.S. is located near Mount Hope Bay, Massachusetts and releases approximately 5 million $m^3$ $day^{-1}$ of thermal effluents, which is typically 7 - 10°C higher than the ambient temperature of the intake water and poses threat to aquatic sustainability (Mustard et al., 1999). The Clean Water Act Section 316a regulates the effluent temperature from power plants; if the inlet water temperatures exceed the allowable limits prescribed by EPA, power production need to curtailed.

Northeastern University

To project stream temperature in the future, we build nonlinear regression models, Support Vector Regression (SVR) using downscaled climate model outputs (at 0.125° spatial resolution). SVR performs better over conventional multiple linear regression and Artificial Neural Network (ANN) based approaches (Shawe-Taylor and Cristianini, 2004). SVR has a simple geometric representation, and it does not depend on the dimensionality of the input data. Although SVR has been applied in several hydrology problems, such as predicting stream flow (flood) (Behzad et al., 2009; Chen and Yu, 2007), soil moisture (Ahmad et al., 2010; Pasolli et al., 2011) and droughts (Chiang and Tsai, 2012; Ganguli and Reddy, 2013), this method has not yet been fully explored in predicting stream temperature.

A summary of the solution framework is shown in Figure 1.1.

## 1.5 Proof of Concept and Lessons Learned

A comprehensive assessment of power production at risk due to freshwater scarcity and increase in stream temperature for the next 30 years will require analysis of data from multiple climate models, multiple initial conditions, and multiple future emissions scenarios. The CMIP5 archive has projected monthly precipitation and evapotranspiration data available from a combination of more than 30 climate models, 4 GHG emissions scenarios, and more than 300 initial conditions run. Each of the model, scenario, and initial conditions combinations gives a possible future climate, and there is no strong basis to include some scenarios and exclude others, especially for the next 30 years when distinguishing between different sources of uncertainties is difficult. Analysis of this huge data set is possible but will require significant amounts of computing and human resources.

In the present work, we have demonstrated how a risk analysis framework can be applied to assess the vulnerability of thermoelectric power plants owing to water stress. To complete the work in the stipulated time frame, we obtained data from three climate models (two U.S. and one Japanese models; U.S. models include the NCAR/DOE model), two initial conditions, and two GHG emissions scenarios (RCP2.6 and RCP8.5; RCP stands for Representative Concentration Pathways). A detailed description about climate data is given in Chapter 2. Data from climate models are available at much coarser spatial resolutions (approximately 110 km); hence, we spatially interpolate both precipitation and evapotranspiration at county

Northeastern University

SDS

levels within ArcGIS. Stream temperature data is not directly available from the CMIP5 archive of climate models. We develop nonlinear regression models based on a combination of downscaled air temperature from climate models and observed water temperature from USGS stream gauges. Subsequently we define two risk metrics - one based on water scarcity and another based on water temperature - and used them to quantify the power production at risk over the mainland U.S.

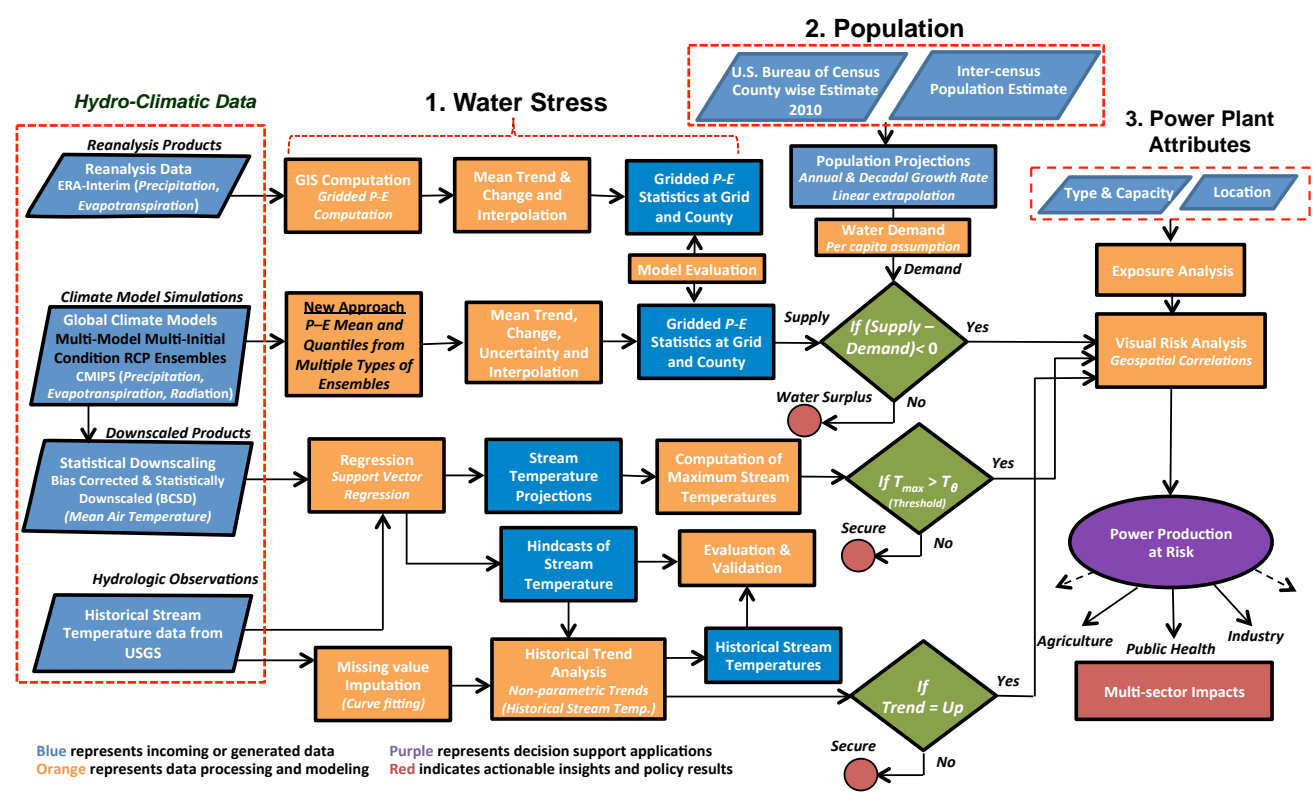

**Figure 1.1** Schematic of solution framework.

## 1.6 Organization of the Report

The rest of the report is organized as follows. The sources and details of climate, hydrologic, population, and power plant data are described in Chapter 2. Methodologies to compute future water supply, projected population, and stream temperature are also described in Chapter 2. We define two metrics to quantify power production at risk and discuss results related to water availability and stream temperature in Chapter 3. In Chapter 4, we discuss which regions of U.S. are most likely will be vulnerable for power production due to water stresses under climate change. Conclusions and discussion are presented in Chapter 5. We discuss challenges and recommendations to overcome them in Chapter 6.

Northeastern University

SDS

# Chapter 2: Data and Methodology

In this chapter, we describe the climate, population, hydrologic, and power plant data along with their sources used in this study in details. We define two metrics to assess power production at risk. In addition, we describe the nonlinear regression models used to project stream temperature.

## 2.1 Data

### 2.1.1 Climate Model Data

*For Estimating Freshwater Availability*

Monthly precipitation and evapotranspiration data were obtained from the latest generation of climate models participating under the Coupled Model Intercomparison Project Phase 5 (CMIP5; Taylor et al., 2012). The latest assessment reports (Stocker et al. 2013: AR5 reports) by the Intergovernmental Panel on Climate Change (IPCC) were compiled from the research performed from the same set of climate models. For our analysis, we extracted data from three global circulation models (GCMs). The name of the models with their modeling group and horizontal grid size is summarized in Table 2.1. We have used two climate models from the U.S. (CCSM4: Community Climate System Model, version four and GISS-E2H: Goddard Institute Space Studies Model E, with HYCOM Ocean Model) and one model from the Japanese (MIROC5: Model for Interdisciplinary Research for Climate, version 5) modeling center. CCSM4 is the U.S. Department of Energy (DOE) model, developed and maintained at the National Center for Atmospheric Research (NCAR). Horizontal grid size of a model represents the number of longitude and latitude (or the uniform spacing between two longitudes or two latitudes) used to discretize the earth to solve the equations of motions of fluids at these grid points. The three models used in the analysis have different grid size (Table 2.1). 1-degree at the equator corresponds to approximately 111 km (~70 miles). The rationale behind using data from more than one model is the difference among them in the numerical modeling of various atmospheric processes; in parameterization schemes to represent clouds; and in the inclusion of different enhanced features such as carbon cycle feedback, dynamic land vegetation, and biogeochemistry. We spatially interpolate precipitation and evapotranspiration from models native grid to a common grid of 2-degree

Northeastern University

SDS

spatial resolution (Diffenbaugh et al., 2013; Giorgi, 2006). The climate data are obtained from the website (http://cmip-pcmdi.llnl.gov/cmip5/) of the Program for Climate Model Diagnosis and Intercomparison (PCMDI).

**Table 2.1** List of CMIP5 models used in the analysis

| Modeling Group | Model Name | Grid Size |
| --- | --- | --- |
| National Center for Atmospheric Research | CCSM4 | 1.25° x 0.94° |
| NASA Goddard Institute for Space Studies | GISS-E2H | 2.5° x 2.0° |
| Atmosphere and Ocean Research Institute (The University of Tokyo), National Institute for Environmental Studies, and Japan Agency for Marine-Earth Science and Technology | MIROC5 | 1.4° x 1.4° |

In the present work, we have extracted projected climate data (2008-2042) from two future scenarios: RCP2.6 and RCP8.5; together they cover the entire range of 21$^{st}$ century radiative forcing scenarios (Rogelj et al., 2012) as shown in Figure 2.1. RCP2.6 represents very low greenhouse concentration level whereas RCP8.5 represents the highest emission scenario. In RCP2.6 scenario, radiative forcing reaches a value of ~ 3.1 W/m$^2$ by mid-of-the-century, and then decreases to ~ 2.6 W/ m$^2$ by 2100 with median global warming of 1.5°C above the preindustrial (Moss et al., 2010). RCP8.5 represents an increase in global radiative forcing of ~ 8.5 W/m$^2$ by the late 21$^{st}$ century, with median global warming of 4.9°C above the preindustrial (Moss et al., 2010).

We have discussed the importance of the role of internal variability in projecting future climate. Initial condition ensembles involve the same model, with same atmospheric physics, run from a different start dates. Presently there is no consensus on how many initial conditions should be used in an analysis; we use data from 2 initial conditions for each model in Table 2.1 to demonstrate as a proof of concept. The objective is to highlight the importance of climate natural variability in the context of water availability analysis, since this has not been shown in earlier literature. In addition, our analysis is constrained by the limited availability of data for multiple initial conditions from the CMIP5 suite of climate models. Only a few selected climate models provide data for more than one initial condition, and even these models do not provide data for more than 2-5 initial conditions. We obtained the data for two initial conditions: *r1i1p1* and *r2i1p1*; the names are just an identifier within the CMIP5 data archive.

Northeastern University

SDS

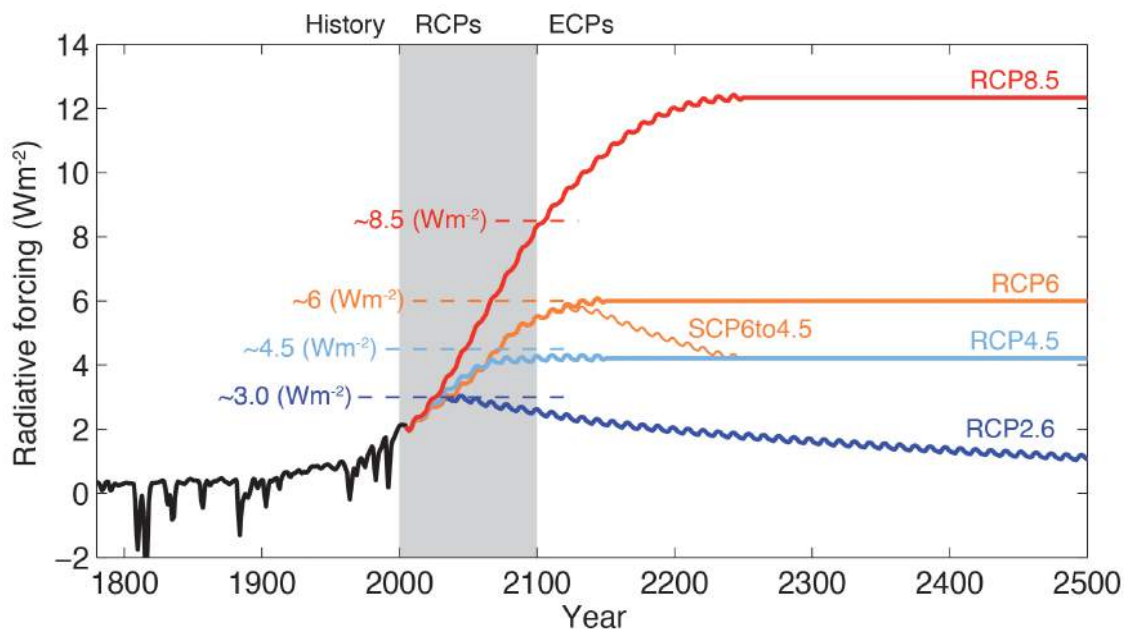

**Figure 2.1** Total RF (anthropogenic plus natural) for RCPs and extended concentration pathways (ECP) – for RCP2.6, RCP4.5, RCP6, and RCP8.5, as well as a supplementary extension RCP6 to RCP4.5 with an adjustment of emissions after 2100 to reach RCP4.5 concentration levels in 2250 and thereafter. (Source: IPCC AR5 Working Group I report, Page No. 147).

*For Projecting Stream Temperature*

The above data set are used to compute projected changes in water supply. Next, we describe the climate data set used to calculate projected water temperature. Presently stream temperature data is not directly available for obtained from the CMIP5 climate models. We develop nonlinear regression models to project water temperature in the future using a combination of climate and hydrologic data. We assume that stream temperature will be related to surface air temperature and longwave and shortwave radiation. We extract the data for these variables from three climate models: CCSM4, MIROC5, and GISS-E2R. The first two models are the same as in Table 2.1, and the third model (GISS-E2R) comes from the same modeling group (NASA) as the model, GISS-E2H. The two NASA models use different ocean models: GISS-E2H uses the HYCOM ocean model and the GISS-E2R uses the Russel ocean model. Further details about the climate models are not necessary to perform the analysis and interpret the results. To build the regression model, we use observed water temperature from the USGS gauge stations. The data from GCMs are at much coarser resolution and are not credible to use as predictors at the spatial resolution of stream gauge

Northeastern University

SDS

location. The standard procedure is to downscale the variable of interests from the GCMs to the smaller spatial scale. The downscaled data for several climate variables from the CMIP5 suite of GCMs exists and have been archived at the following website: http://gdo-dcp.ucllnl.org/downscaled_cmip_projections. The archived data uses Bias-Corrected Statistical Downscaling (BCSD) methodology for the downscaling. We obtained mean air temperature data at the spatial resolution of 0.125-degree (~9 miles) for the three climate models: 5 initial conditions run for CCSM4 and one initial condition run for each MIROC5 and GISS-E2R. In our initial exploration phase to identify a set of predictors for the regression model, we also used longwave and shortwave radiation as potential predictors. The downscaled data for these variables is not archived, so the data is directly obtained from CMIP5 archive. Please refer to the method section for further details. In the next section, we describe the hydrological variables.

### 2.1.2 Hydrologic Data

Historical stream temperature data of 18 major hydrologic units (with 332 gauge stations) at monthly time scales are obtained from the USGS web site (Source: http://waterdata.usgs.gov/nwis) based on data availability. Stream temperature data with at least 7-years of record are selected for historical trend analysis. Less than 7-years data may not be sufficient for estimating trends due to seasonality effect. Data availability from different gauges varies between the years 1969 to 2012. Figure 2.2 show spatial maps of hydrologic unit and stream gauge locations therein. The names of major water resources regions in the conterminous United States are summarized in Table A3.

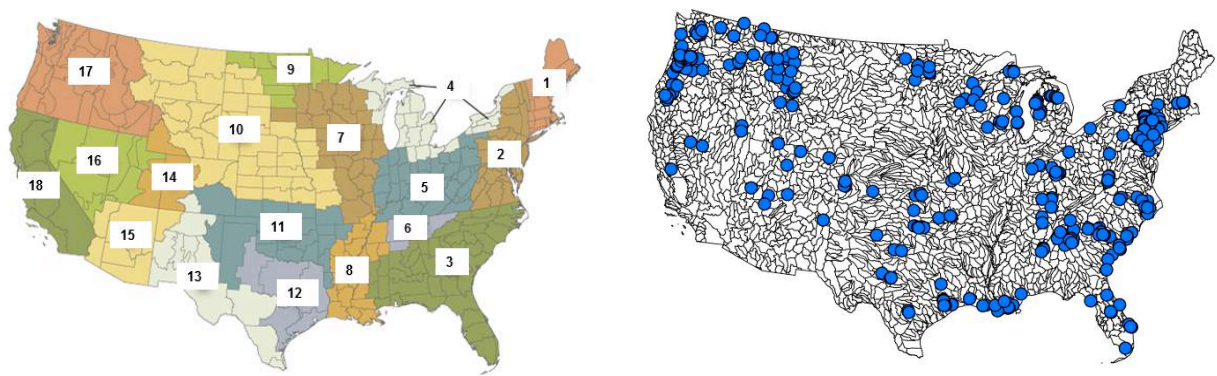

**Figure 2.2** (*left*) Major water resources region in Conterminous United States and (*right*) spatial locations of 332 stream gauges (shown in blue solid circles) over water resources region.
(Source: http://www.ncdc.noaa.gov/)

Northeastern University

SDS

### 2.1.3 Population Data

To estimate future water demand for municipal and domestic supply at county levels, we need projected population data in the respective years (2030s and 2040s). The projected population data is available from the U.S. Census Bureau at national level but not at county levels. We followed the procedure described in Roy et al. 2012 to make population projections in 2030 and 2040 at county levels. The expressions to compute the future population are given below. To estimate county level population projection, county level data for the period 2000-2010 from U.S. Census Bureau are used to compute annual growth rate for each county in percent per year. (https://www.census.gov/topics/population.html). The spatial distributions of population at county levels for the year 2010 are shown in Figure 2.3.

The projected populations in 2030 and 2040 are calculated as follows (Roy et al., 2012):

- Estimate mean annual population growth rate for each county from historical population data (2000-2010)
- Population in 2030 (county) = population in 2010 × (1 + mean annual growth rate)^20
- Population in 2040 (county) = population in 2010 × (1 + mean annual growth rate)^30

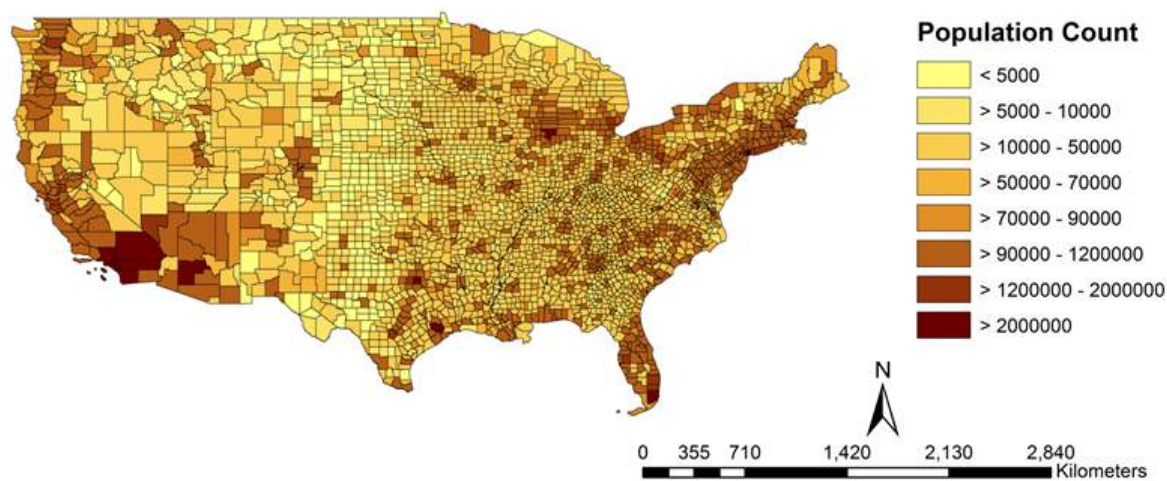

**Figure 2.3** spatial distributions of county level population data from the U.S. Census Bureau

We evaluated the robustness of future estimates of population by aggregating the county level projected population at national level and then compared with the projected data available from the U.S. Census Bureau. The percentage differences between the aggregated population at national level by the methodology suggested above and that from the projected data given

Northeastern University

by the U.S. Census Bureau are 2.6% and 8.4% in 2030 and 2040, respectively. The projected population at the county level in 2030 and 2040 is shown in Figure 2.4. Dense population is observed in Northeast, Florida, Pacific Northwest, and coastal Southwest.

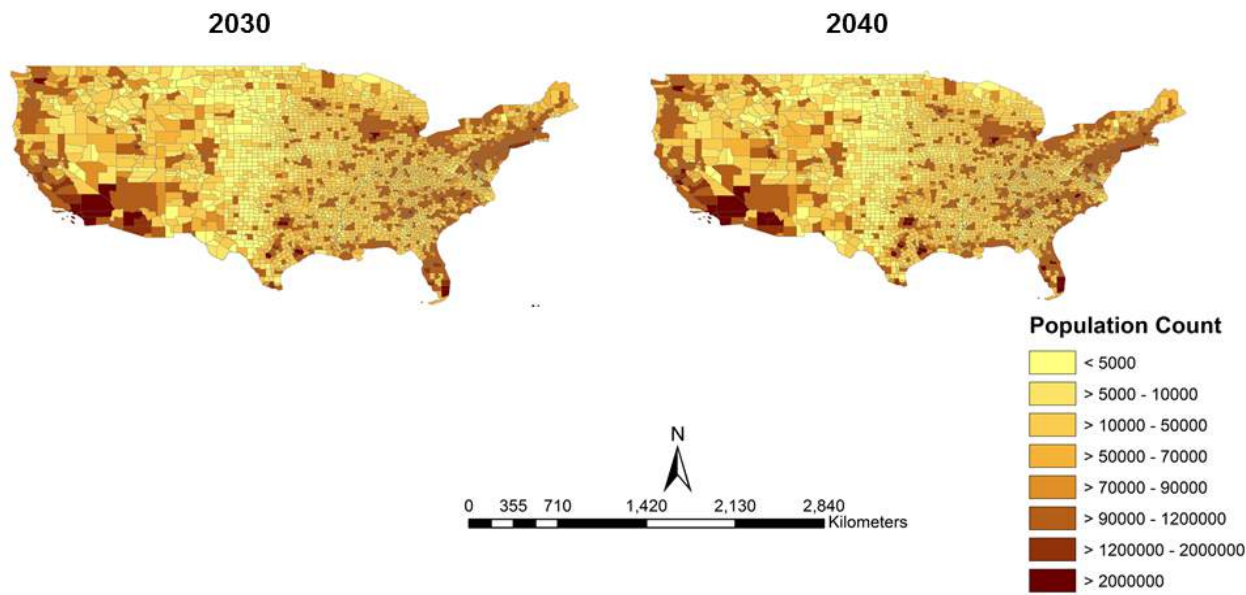

**Figure 2.4** Spatial distributions of projected population at county level in 2030 and 2040

Gridded projected population data at varying spatial resolutions are available from other sources; one of the most commonly used are the Gridded Population of the World (GPWv3) from the Center for International Earth Science Information Network (CIESEN), Columbia University. We did not use these data sets for the following reasons:

- Aggregation of the projected population data from grid to county scale results in zero human count in some counties
- Disaggregation methodology to distribute population at fine spatial resolutions results in fractional human count.
- In most of the gridded population database, the projected population has been computed by assuming 2000 as the base year

### 2.1.4 Power Plant Data

Spatial locations of thermoelectric power plants along with their capacities were compiled primarily from two sources: the Electric Power Research Institute (EPRI, 2011) and the Energy Information Administration database (EIA, 2013). In these data sources, power plant cooling systems have been divided into 4 categories:

1. **Direct or once-through:** This kind of cooling system takes water from nearby sources (such as rivers, lakes, reservoirs, or the ocean), circulates it through pipes in a single cycle to absorb the surplus heat from steam, and then discharges the warmer water into the environment. This type of cooling system withdraws large amounts of water and has greater potential to harm the local ecosystems compared to other cooling systems.

2. **Wet-recirculating or closed-loop:** This system withdraws water only to replace the lost to evaporation and is used primarily in regions where water resources are not abundant. The wet-recirculating systems have much lower water withdrawals than once-through systems; however, they have higher water consumption.

3. **Dry-cooling:** Such systems use air instead of water to absorb the surplus heat energy exiting the turbines. Dry-cooled systems can reduce the water consumption by more than 90% compared to wet-recirculating systems. Although no water is required for cooling but it is required for maintenance and cleaning.

4. **Hybrid-Cooling**: Combined cooling systems have both wet as well as dry cooling components. Hybrid cooling uses both air and water for cooling and consumes 50% less water than a conventional closed-loop wet cooling system (Feeley et al., 2006)

In this report, we consider once-through and wet-recirculating cooling systems collectively as wet cooled systems (and thus, subject to water vulnerability). The nomenclature has been used in the analysis and in figures to show the results. Figure 2.5 shows capacity (in Gigawatts) of thermoelectric power plants by types of cooling systems and fuel used to generate steam in 2010. In the U.S., 98% of thermoelectric power plants fueled by coal, nuclear, natural gas, and other sources use water for cooling (U.S. DOE/EPSA-0002, 2014; EIA, 2013).

Northeastern University

SDS

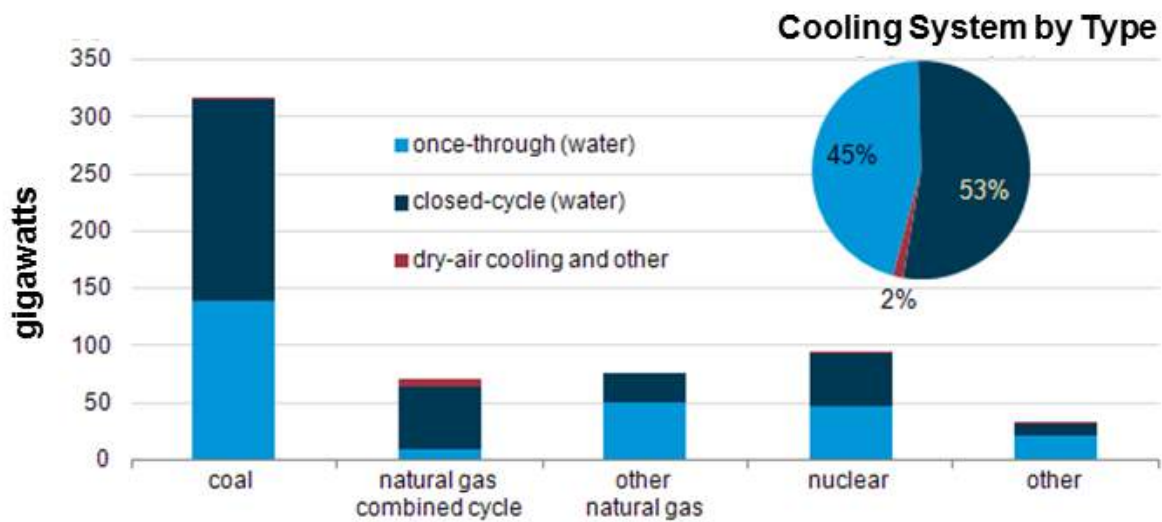

**Figure 2.5** Capacity of thermoelectric power plants by cooling system type and fuel, 2010. (Source: http://www.eia.gov/todayinenergy/)

The spatial distributions of power plants with the cooling methods (Wet cooled versus others) employed and their electricity generation capacities, in Quadrillion British Thermal Units (QBTU[3]; in short Quad), are shown in Figure 2.6.

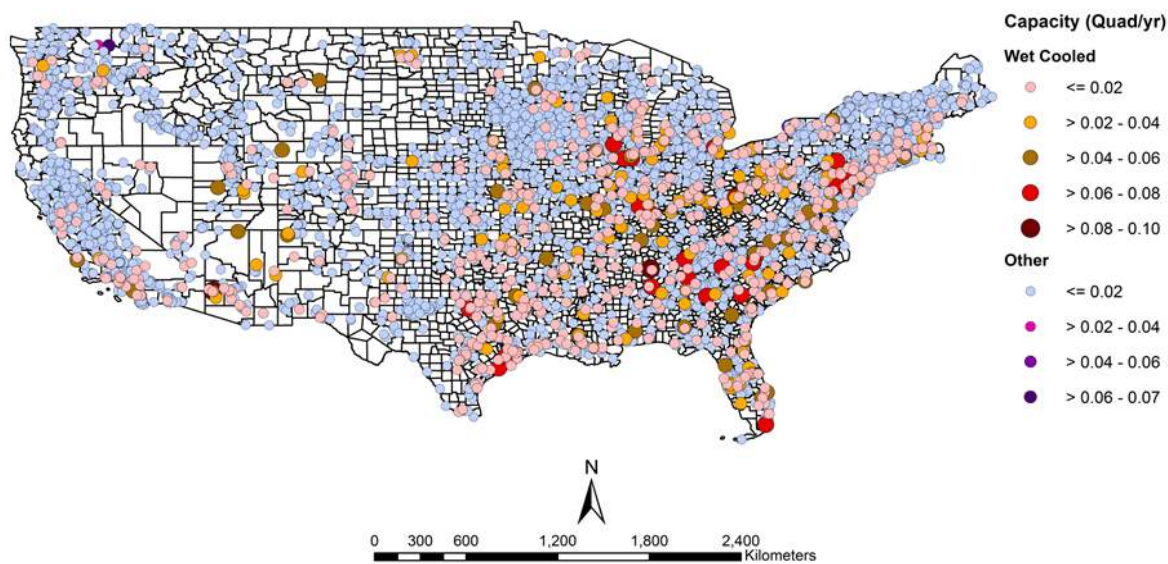

**Figure 2.6** Location of wet cooled and other thermoelectric power plants along with their electricity generation capacities (Quad/year)

---

[3] Quadrillion British thermal units is a common metric to describe energy use across all energy resources. A British thermal unit is equal to 1,055 joules. A single QBTU would provide all of the energy demand for New York State for approximately three months. [Source: Resource Revolution: Meeting the world's energy, materials, food, and water needs, Nov 2011, the McKinsey Global Institute]

**Northeastern University**　　　　　　　　　　　　　　　　SDS

As mentioned above, we have also analyzed the impact of projected stream temperature on the amount of energy production at risk. The future water temperature is projected using nonlinear regression model based on the historical stream temperature and other climate variables. Therefore, it is important to collect the stream temperature data close to the power plant location. In this analysis, we have taken the stream temperature record from those USGS locations that are in close proximity to the power plants.

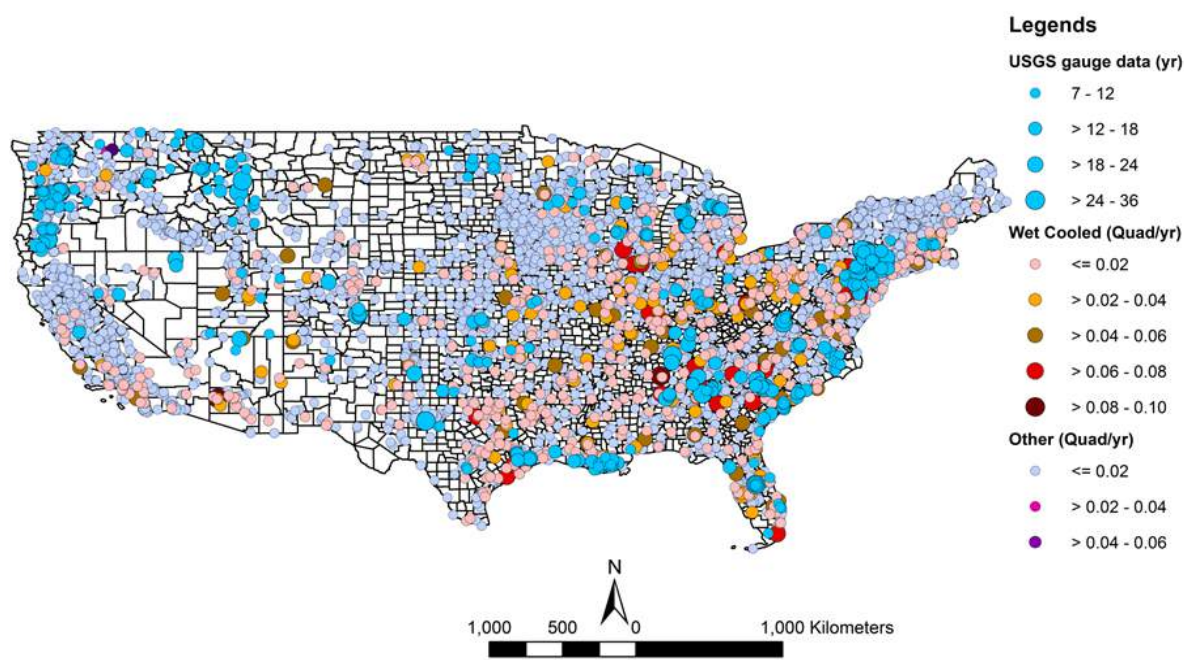

**Figure 2.7** Location of wet cooled and other thermoelectric power plants along their capacity and the location of USGS stream gages.

We overlay the locations of stream gages in Figure 2.7 (the size of filled blue circles shows the length of data available) to show their proximity to the power plants (the red filled circles in different size and shade show the installed capacity of power plants) as shown in Figure 2.6. The wet cooled power plants produce the maximum amount of energy in the U.S.; we focus our subsequent analysis on the wet cooled plants. In Figure 2.8, we have shown the location of wet cooled plants (red filled circles with size indicating their installed capacity) only and the corresponding locations of USGS stream gauges (blue filled circle).

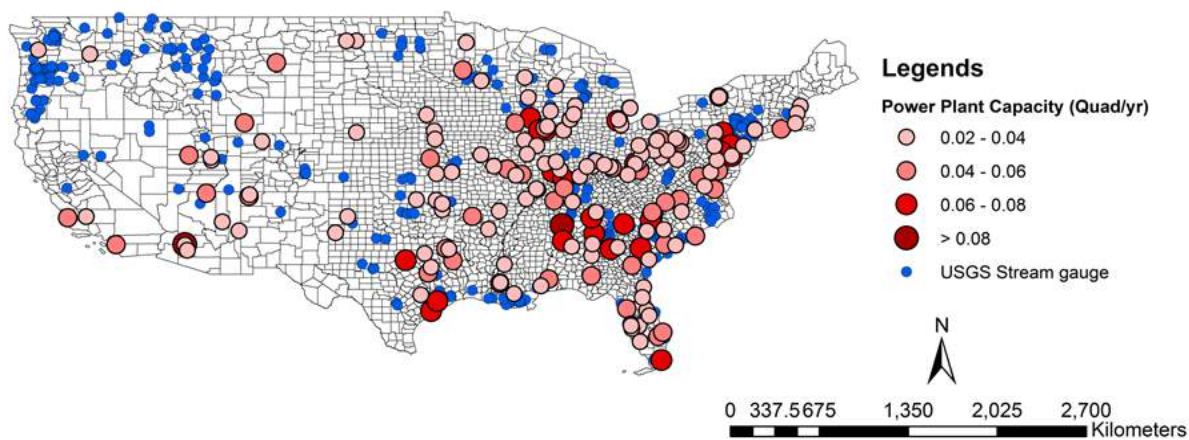

**Figure 2.8** Spatial distributions of wet cooled power plants and of USGS stream gauge locations

## 2.2 Methods

### 2.2.1 Metrics for Estimating Power Production at Risks

In consultation with ARPA-E, we defined two metrics – Water Availability Absolute Change Index (WAACI) and Water Temperature Stress Index – to quantify the amount of power production at risk.

#### *Water Availability Absolute Change Index (WAACI)*

A first order estimate of water availability is calculated by taking the difference between precipitation and evapotranspiration. Contributions from groundwater and inter-basin transfers are not considered in the first-order estimate. Precipitation from the climate models includes contribution from rain and snow. Net water availability is defined as the difference between water supply and water demand from all sectors. For the present work, we have assumed that future water demand other than municipal public and domestic supply and thermoelectric cooling will not change. The bases for assumptions have been discussed above. Future water supply is estimated by taking the climatological mean of surface runoff (precipitation minus evapotranspiration) to include the effect of climate change and variability. Future municipal and domestic water demand is estimated using the projected population and per capita water use, which we have taken to be 1700 m$^3$/capita/year (Parish et al., 2012; Falkenmark, 1986). Thus future change in water demand is computed as the product of change in population and water demand per capita (1700 m$^3$/capita/year). WAACI values are computed at each grid point as the difference between water supply and water

**Northeastern University**          **SDS**

demand as follows:

$$WAACI = \left[\frac{1}{n} \times \sum_{i=1}^{n} \left(P - E\right)_t\right] - per\ capita\ water\ demand \times population \quad t = 1,...,n \qquad (1)$$

Where $t$ and $n$ denote changes in available freshwater (*i.e.*, $P - E$) averaged over $n$ duration (years) at $i^{th}$ time steps (in months). The gridded estimates of WAACI are spatially interpolated within ArcGIS to get the values at the county level.

### *Water Temperature Stress Index*

We consider a power plant in the vicinity of a stream gauge is under stress when the maximum stream temperature exceeds EPA allowable limit, which is defined as,

$$T_{WTSI} = 1 \times \{T_{stream} > T_{EPA}\} \qquad (2)$$

Where $1\{\Phi\}$ is a logical indicator function of set $\Phi$ that takes the value of either 0 (if $\Phi$ is false) or 1 (if $\Phi$ is true). The value of $T_{WTSI}$ is either 0 or 1. Table A4 outlines allowable limits for stream temperature according to EPA regulations for different states.

### 2.2.2 Analysis of Trend in Historical Stream Temperature

As stated above, the projected water temperature is not available directly from CMIP5 climate models. We develop nonlinear regression models to predict water temperature in the future using a combination of climate and hydrologic variables as predictors. Our first objective is to find a set of predictors that will give us reliable projections of stream temperature in the 2030s and 2040s. To that end, we also examine whether we should include lagged variables in the set of predictors to include seasonality effects. We perform a trend analysis on the observed stream temperature data for all stations where the length of the recoded data is at least 7 years. The missing data in the observed record are imputed using a time series interpolation technique, which is based on sparse linear algebra and PDE discretization (D'Errico, 2004).

The illustration of the method to impute missing values is shown in Figure 2.9 for the USGS stream gauge at Lees Ferry at Colorado River in the state of Arizona. Serially complete

Northeastern University

SDS

stream temperature time series without any missing data is required for trend analysis. We perform nonparametric trend analysis using the Mann-Kendall test with correction for ties and autocorrelation (Hamed and Ramachandra Rao, 1998; Helsel and Hirsch, 1992). We accounted for autocorrelation in trend analysis because water temperature time series exhibit strong seasonality.

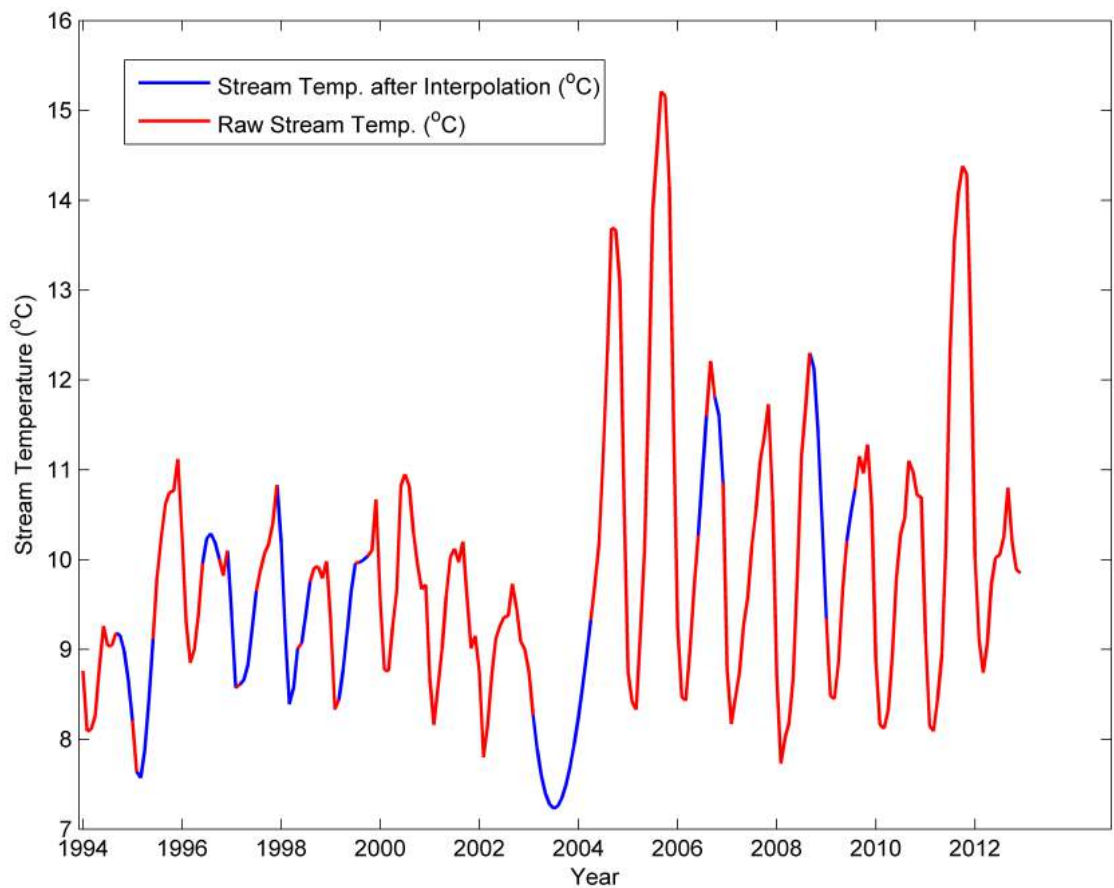

**Figure 2.9** Time series of historical stream temperature data after filling missing values a station Lees Ferry on the Colorado River in the state of Arizona

To illustrate seasonality, we show an autocorrelation function for water temperature at one of the stream gauge location in Georgia in Figure 2.10. The figure suggests inclusion of lagged variables into the regression model. We develop Support Vector Regression (SVR) models with multiple sets of predictors and identify a set of the predictors that will give a credible projection of stream temperature at a monthly time scale. The computation is carried out in the commercially available software MATLAB using the StatLSSVM package (De Brabanter et al., 2013). We started with the following set of predictors obtained from the climate models:

Northeastern University

SDS

- Downscaled mean air temperature ($t_{air}$) at three times: *t, t-1, and t-2*
- Surface downwelling clear-sky longwave radiation (*rldscs*) at two times: *t* and *t-1*
- Surface downwelling clear-sky shortwave radiation (*rsdscs*) at two times: *t* and *t-1*

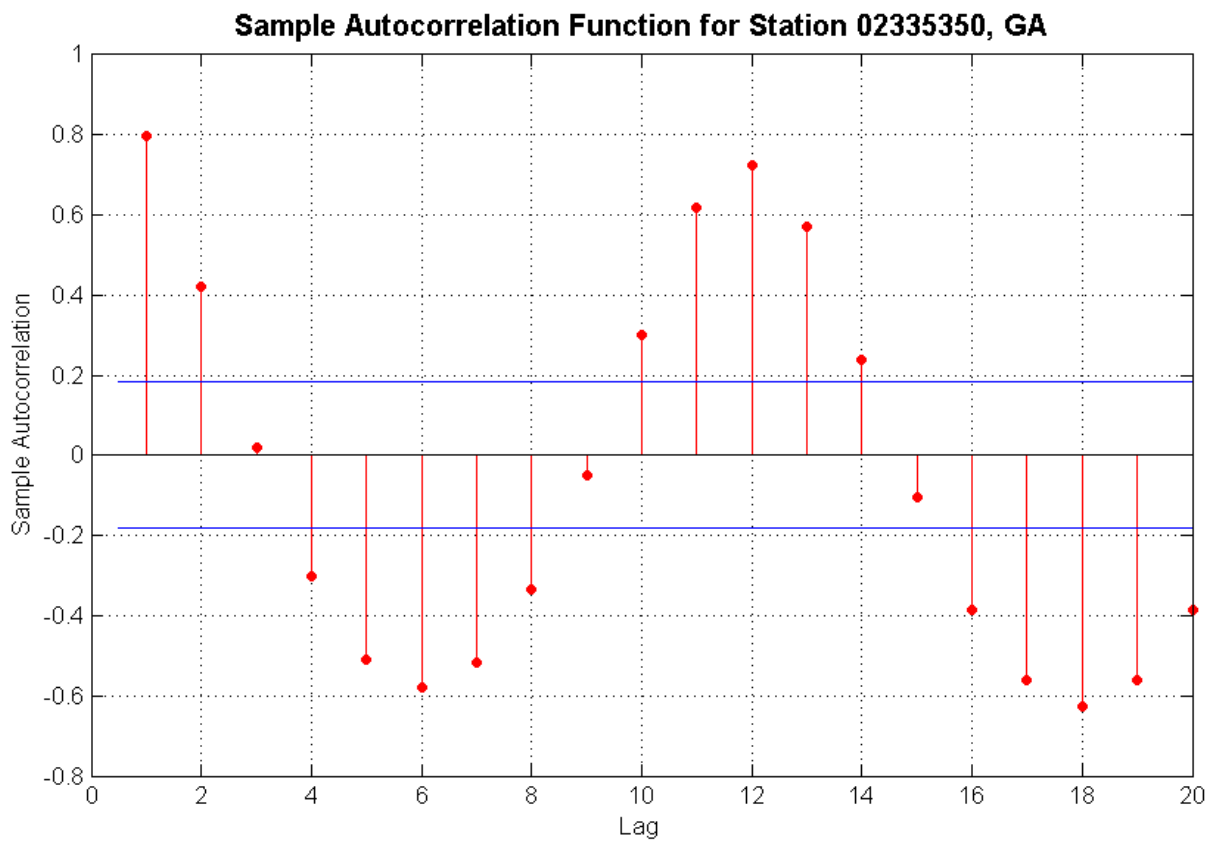

**Figure 2.10** Sample autocorrelation function for the historical stream temperature for USGS stream gauge location 02335350 in the state of Georgia. Blue lines show the autocorrelation of stream temperature with 5% significance limit. The spikes above (below) these limits show autocorrelation is significant.

We have taken predictors from multiple climate models and multiple initial conditions. Each of the combination presents a plausible future climate, and a rigorous analysis should include the projection of water temperature for all of them separately. However, in this proof of concept analysis, we have performed the analysis for multimodel ensemble median and multimodel ensemble second maxima. A functional relationship between the projected stream temperature and lagged input variables can be written as

$$T_{stream\ projected}\ (t+l) = f\ (T_{stream\ projected}\ (t-z_1),\ rldscs\ (t-z_2),\ rsdscs\ (t-z_3)) \qquad (3)$$

Northeastern University

SDS

Where, $l$ is the forecast lead time; $z_1$, $z_2$, $z_3$ are the time lags considered for the predictors taking a value of 0, 1 and 2 months, respectively; $t$ denotes the current time step of historical stream temperature. We use 10 years of data (1998 – 2007) to train the model and 5 years of data (2008 – 2012) to validate the model. The performance of different sets of predictors is evaluated using the Nash-Sutcliffe Efficiency (*NSE*) Index and Pearson's linear correlation (*r*). Table 2.2 shows SVR-based model performance for stream temperature during the training and validation phase using a multimodel median of climate models as predictor, at Colorado River, Lee's ferry. Model 4 performs best and hence, for the subsequent analyses we use current and lagged air temperatures as predictors.

**Table 2.2** The performance of different SVR models.

| Models | Predictors | Training | | Testing | |
|---|---|---|---|---|---|
| | | *NSE* | *r* | *NSE* | *r* |
| Model - 1 | $t_{air}(t)$, $t_{air}(t$-1$)$, $t_{air}(t$-2$)$, $rldscs(t)$, $rldscs(t$-1$)$, $rsdscs(t)$, $rsdscs(t$-1$)$ | 0.505 | 0.712 | 0.252 | 0.601 |
| Model – 2 | $t_{air}(t)$, $rldscs(t)$, $rsdscs(t)$ | 0.434 | 0.659 | 0.404 | 0.680 |
| Model – 3 | $t_{air}(t)$, $t_{air}(t$-1$)$, $rldscs(t)$ | 0.430 | 0.656 | 0.288 | 0.608 |
| Model – 4 | $t_{air}(t)$, $t_{air}(t$-1$)$, $t_{air}(t$-2$)$ | 0.392 | 0.644 | **0.511** | **0.813** |

*Note: NSE* $= 1 - \sum_{i=1}^{n}\left(O_i - P_i\right)^2 \Big/ \sum_{i=1}^{n}\left(O_i - \bar{O}_i\right)^2$, $NSE \in \left(-\infty, 1\right]$, Where $O$ = Observed, $P$ = Predicted values, *NSE* < 0, indicate residual variance (numerator) is greater than data variance (denominator); ensemble median of climate models (with one initial condition runs only) are chosen as predictor; model with best performance is marked in bold.

A plot of observed stream temperature, current and lagged air temperatures at the validation phase for the USGS station at Lee's ferry in Colorado River basin is shown in Figure 2.11. The standardized time series of stream temperature closely follow the ambient air temperature at different time lags except for the year 2011. The abrupt rise in stream temperature in the year 2012 may be attributed to exogenous factor, which is not fully captured by air temperature. Figure 2.12 shows the time series of observed and predicted stream temperatures at different USGS stream gauge locations during the validation phase. A visual examination confirms the satisfactory performance of the regression model.

Northeastern University

SDS

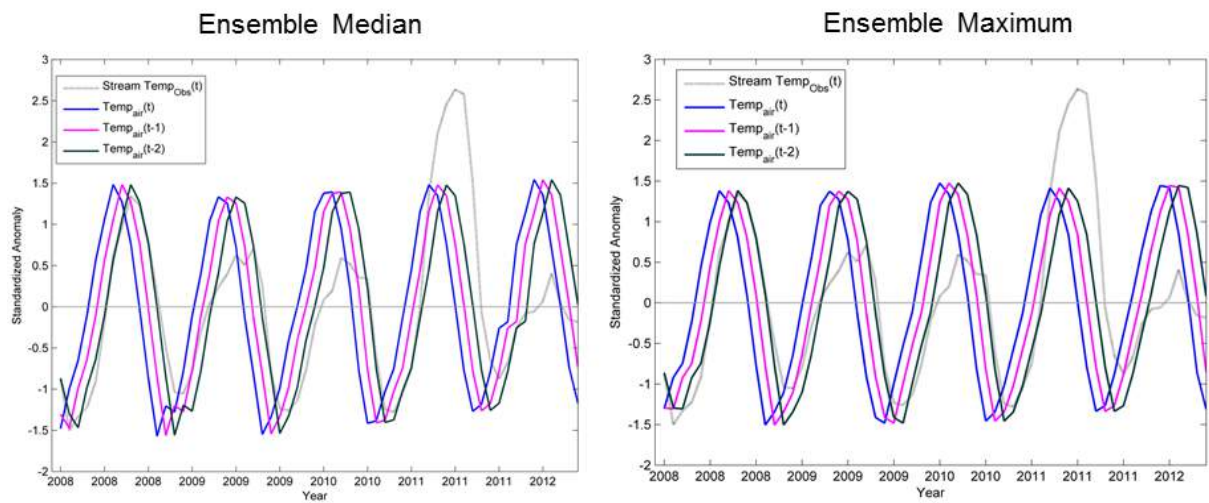

**Figure 2.11** Observed stream temperature, current and lagged air temperatures during validation phase for the Lee's ferry station on the Colorado River

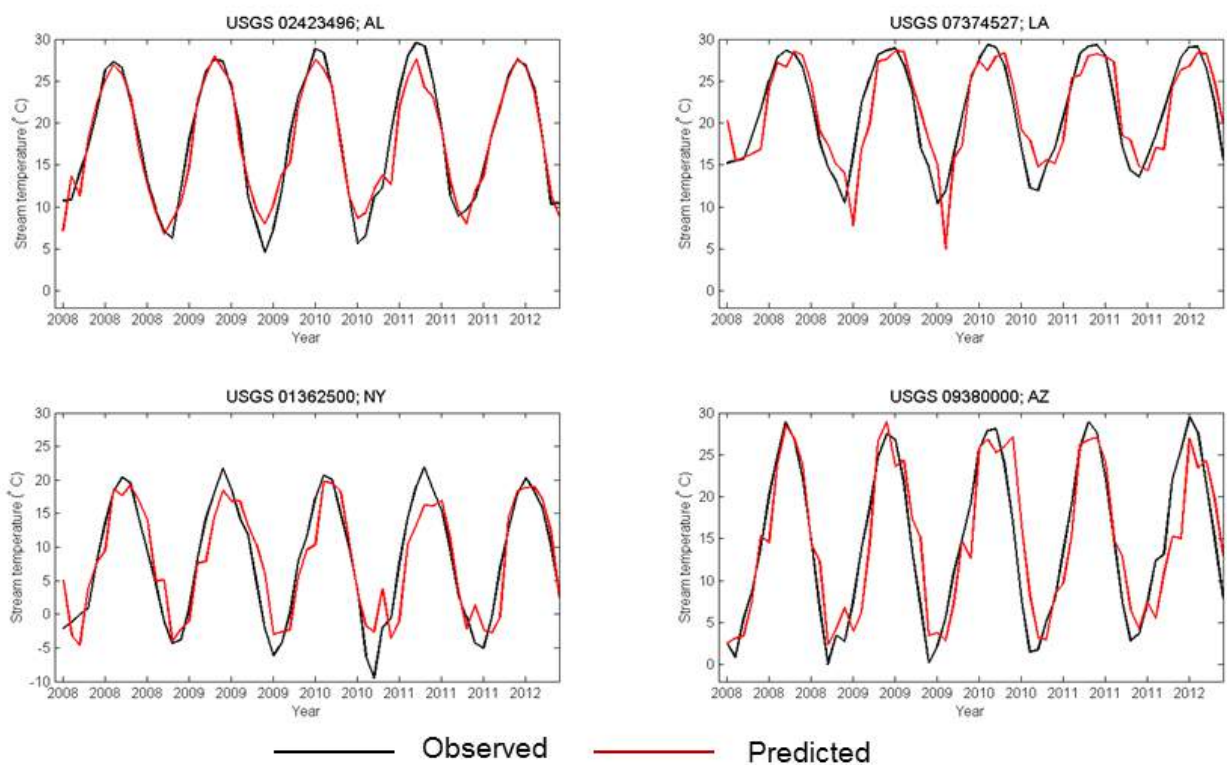

**Figure 2.12** Validation results of stream temperature simulation for different gauges. Ensemble median of climate models are used as predictors for SVR-based regression model

We evaluated the performance of multimodel ensemble (MME) median and MME maximum (ensemble 2nd maximum as defined by the 80th percentile values from all combinations of climate models and initial conditions for a given emission scenario) by plotting a histogram of the Nash-Sutcliffe Efficiency index as shown in Figure 2.13. The class frequency of histogram close to 1 (for Pearson's correlation) is around 15 for MME median as compared to 11 for MME maximum. In terms of predictive skills, the multimodel median performed better; hence, for subsequent analyses we use MME median as predictors for developing regression models.

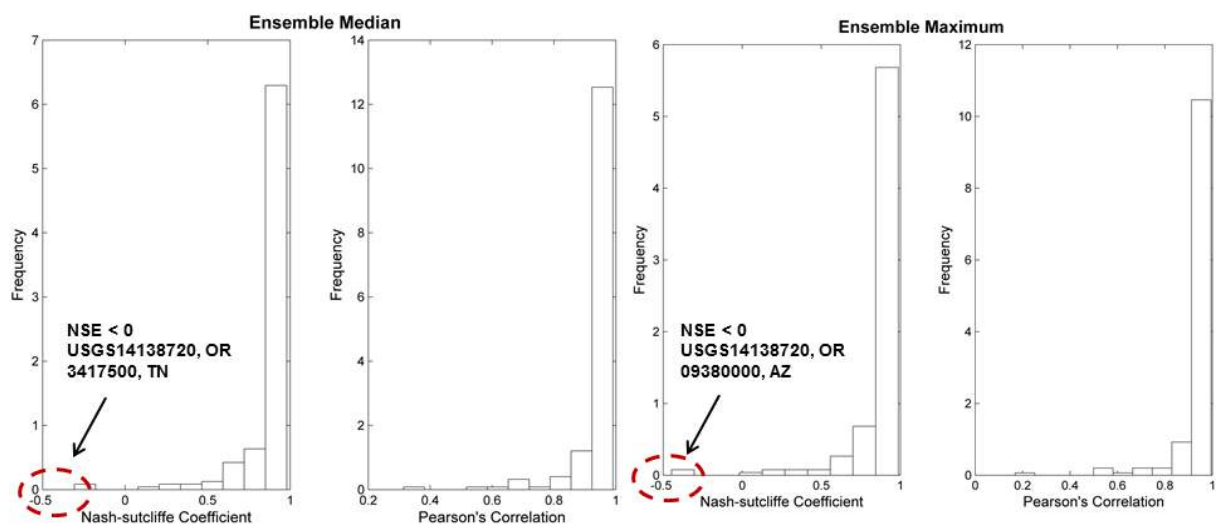

**Figure 2.13** Histograms of performance of SVR-based models in 184 stream gauges during validation phase (2008 – 2012) analyzed by different performance metrics.

Figure 2.14 shows the spatial distribution of performance of the SVR-based models in predicting maximum stream temperature and associated bias in the validation phase. We assume that biases in downscaled GCM output during the future period are similar to that of the historical period (2008 – 2012). Negative bias in stream temperature is observed in the Pacific Northwest, and few scattered locations over the Great plains and Southeast U.S., indicating underestimation of maximum stream temperature by the model. However, positive bias is noted in most of the locations. The maximum negative bias (-1.9°C) in stream temperature is observed in USGS stream gauge location at Deer Lodge county, Montana whereas the maximum positive bias (9.5°C) is noted in the stream gauge at Skamania county, Washington.

Northeastern University

SDS

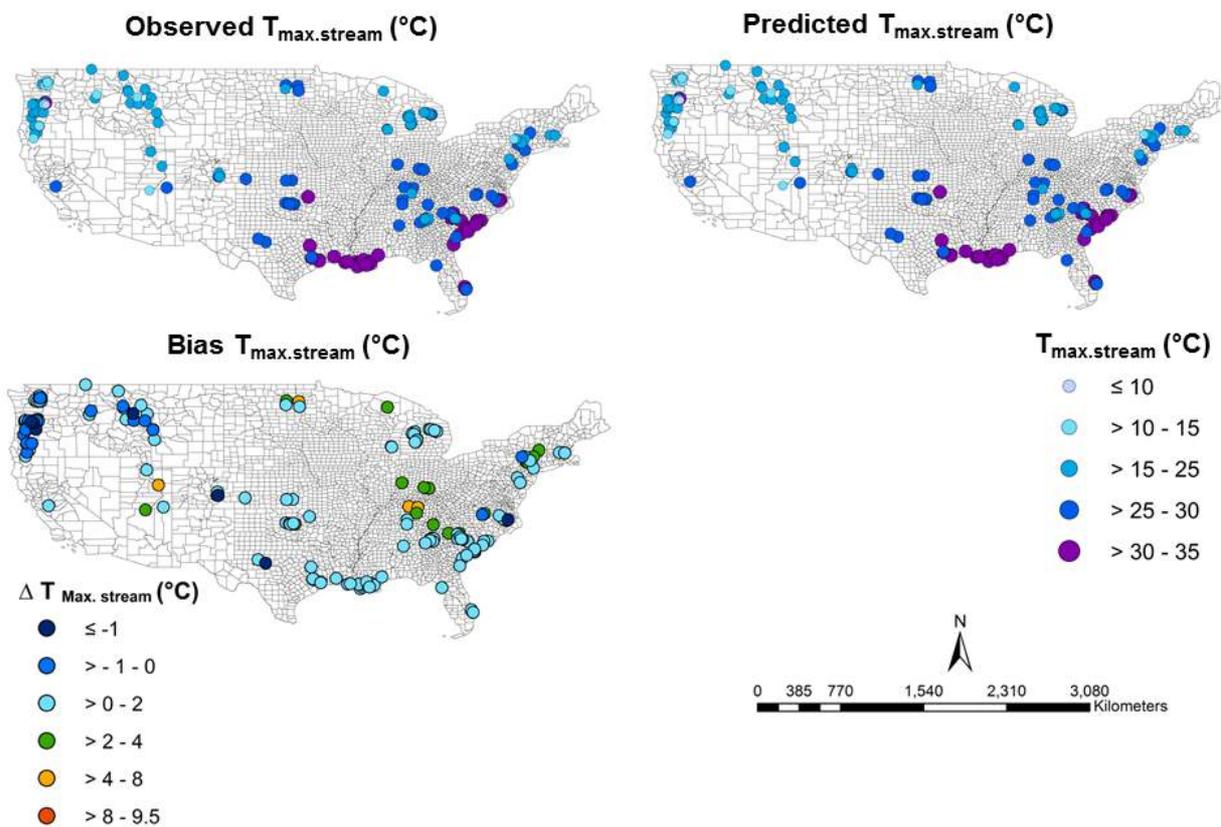

**Figure 2.14** Observed versus predicted maximum stream temperature and the associated bias (predicted – observed) during validation phase. The biases are taken into account for projecting stream temperature.

# Chapter 3: Future Water Availability and Temperature

## 3.1 Decrease in Water Availability

### 3.1.1 Performance of CMIP5 models in simulating current freshwater availability

First, we evaluate the performance of three CMIP5 GCMs considered in the analysis to assess how well they are able to simulate the spatial patterns of freshwater (*P-E*) against observed estimates of available freshwater. Over the U.S., gridded observed precipitation data exists but there are no reliable observed evapotranspiration datasets. When observed data is unavailable, to evaluate the performance of the models against past climatology we use reanalysis[4] data, which are often used as proxies for real observation.

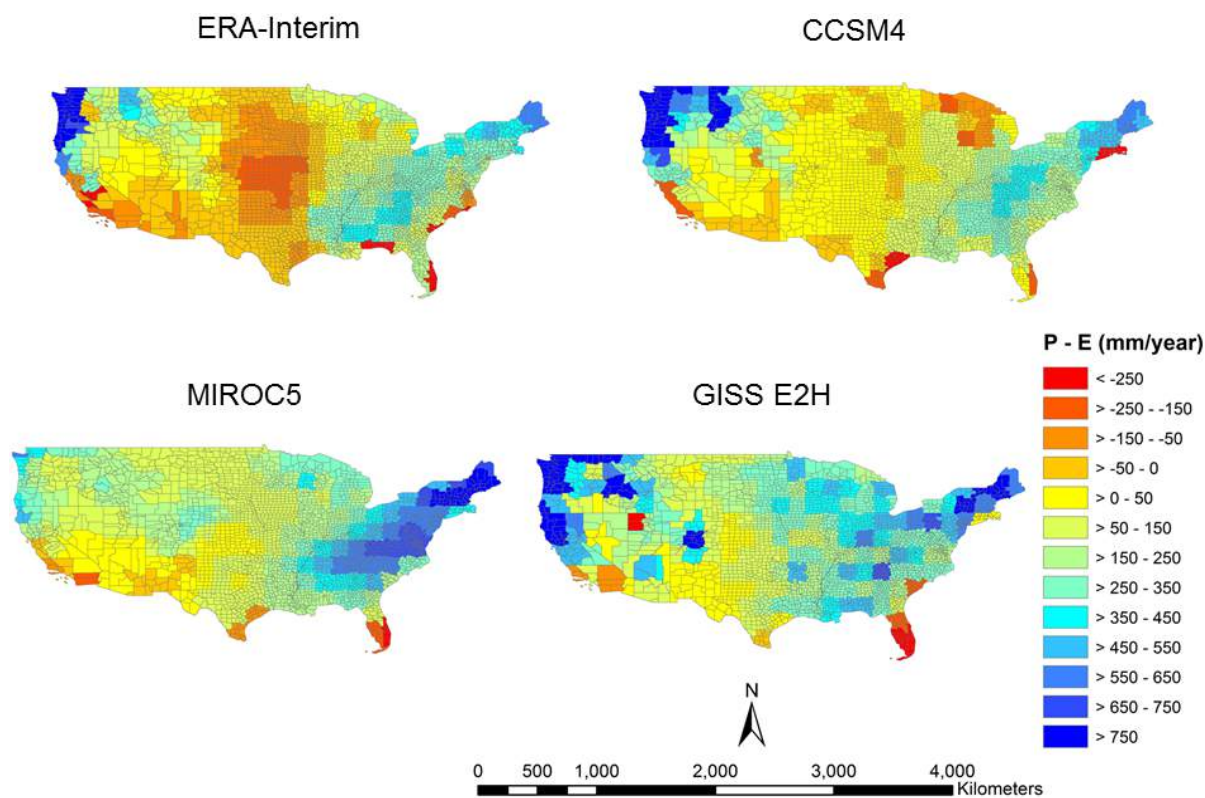

**Figure 3.1** Spatial patterns of freshwater from climate models and ERA-Interim for 2010s

Here we have considered the third generation (the latest generation) of reanalysis products, the observationally constrained European Center for Medium Range Weather Forecasts

---

[4] Reanalysis datasets are created by assimilating ("inputting") climate observations using the same climate model throughout the entire analysis period in order to reduce the effects of modeling changes on climate statistics (Source: http://www.esrl.noaa.gov/psd/data/gridded/reanalysis/)

Northeastern University

SDS

(ECMWF), ERA-Interim (Dee et al., 2011) as the reference dataset for comparing model past performance. Current freshwater availability is computed by taking the five year climatological average (2008-2012) of *P-E* at each grid point from ERA-Interim reanalysis data and for one initial condition (r1i1p1) from three climate models, and subsequently spatially interpolated using ArcGIS to get the estimates at county levels. The spatial patterns of freshwater availability in the 2010s (2008-2012) from climate models and ERA-Interim are shown in Figure 3.1. The reanalysis data depicts drying patterns over the Midwest and wet patterns over small regions of the Pacific Northwest. There are many regions over which spatial patterns of freshwater differ in ERA-Interim and climate models. Inter-model differences in estimates of fresh water over some regions, such as the relatively wet pattern in the Northeast U.S. as simulated by MIROC5 model, can be observed in Figure 3.1.

### 3.1.2 Water Availability Absolute Change Index (WAACI)

We quantify water stress using the Water Availability Absolute Change Index (WAACI). WAACI is computed at each grid point and then spatially interpolated using ArcGIS; this index considers both water supply from climate models and water demand only from municipal supply. WACCI has been computed for multiple combinations of climate models and initial conditions, but here we are showing it for two cases: MME minimum (the 2[nd] minimum of all possible combinations of climate models and initial conditions) and MME median. Figure 3.2 shows spatial distributions of WAACI in 2010s (2008-2012) for MME minimum and MME median of climate models.

We have considered precipitation and evaporation data from climate models under RCP8.5 GHG emissions scenarios. MME minimum projects a drying trend over most of the regions, while MME median shows drying patterns over the Southern part (most part of Texas and Oklahoma), parts of Florida and Southwest regions. However, in both cases, significant dry conditions (WAACI ≤ -3000000 Mgal/year) are evident in coastal California. Water stress is further exacerbated by the increase in population in California, with Los Angeles County showing the highest population (more than nine million) during 2010.

Northeastern University

SDS

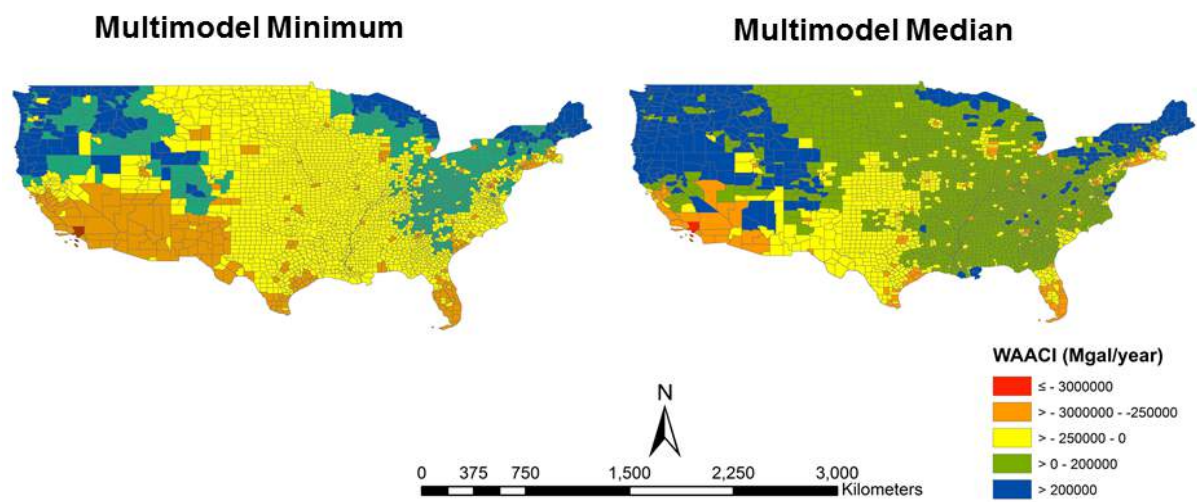

**Figure 3.2** Water Availability Absolute Change Index (WAACI) during 2010s under two cases (i) multi-model minimum and (ii) multi-model median ensemble

### 3.1.3 Uncertainty in Estimation of Water Availability

*Uncertainty among Climate Models and Initial Condition Runs in Estimate of P - E*

In Figure 3.1, we observed inter-model differences in spatial patterns of freshwater in the 2010s with just three models and one initial condition. This motivates us further to look at the uncertainty in the estimate of fresh water from multiple climate models and multiple initial conditions for several time periods in the future. There are many ways in which we can quantify the uncertainty; here we limit ourselves to qualitative comparison by visual inspection of the spatial maps for different models and the corresponding initial condition runs.

Figure 3.3 shows the spatial variability of annual mean changes in freshwater availability from three climate models (CCSM4, GISS-E2H, & MIROC5) and two initial condition runs (r1i1p1 and r2i1p1) over projected 5-year segments (2020s: 2018 – 2022, 2030s: 2028 – 2032, 2040s: 2038 – 2042) for RCP8.5 GHG scenario. In Figure 3.3, we have shown changes in fresh water (*P-E*) with respect to past estimates (2010s: 2008-2012); i.e., for 2020s, we computed the changes in *P-E* at each grid point by taking the difference in 5-year average of *P-E* from 2020s and 2010s. There are multiple ways we can look at the results. For given initial condition (such as *r2i1p1*) and for one time period (2040s), we can see the inter-model differences in the spatial variability of P-E over many regions. In fact, the differences across

Northeastern University                                                    SDS

models and initial conditions are so contrasting that over same regions we have both drying and wet patterns. This propelled further research to question which model and initial condition combination should be used to assess the power production at risk.

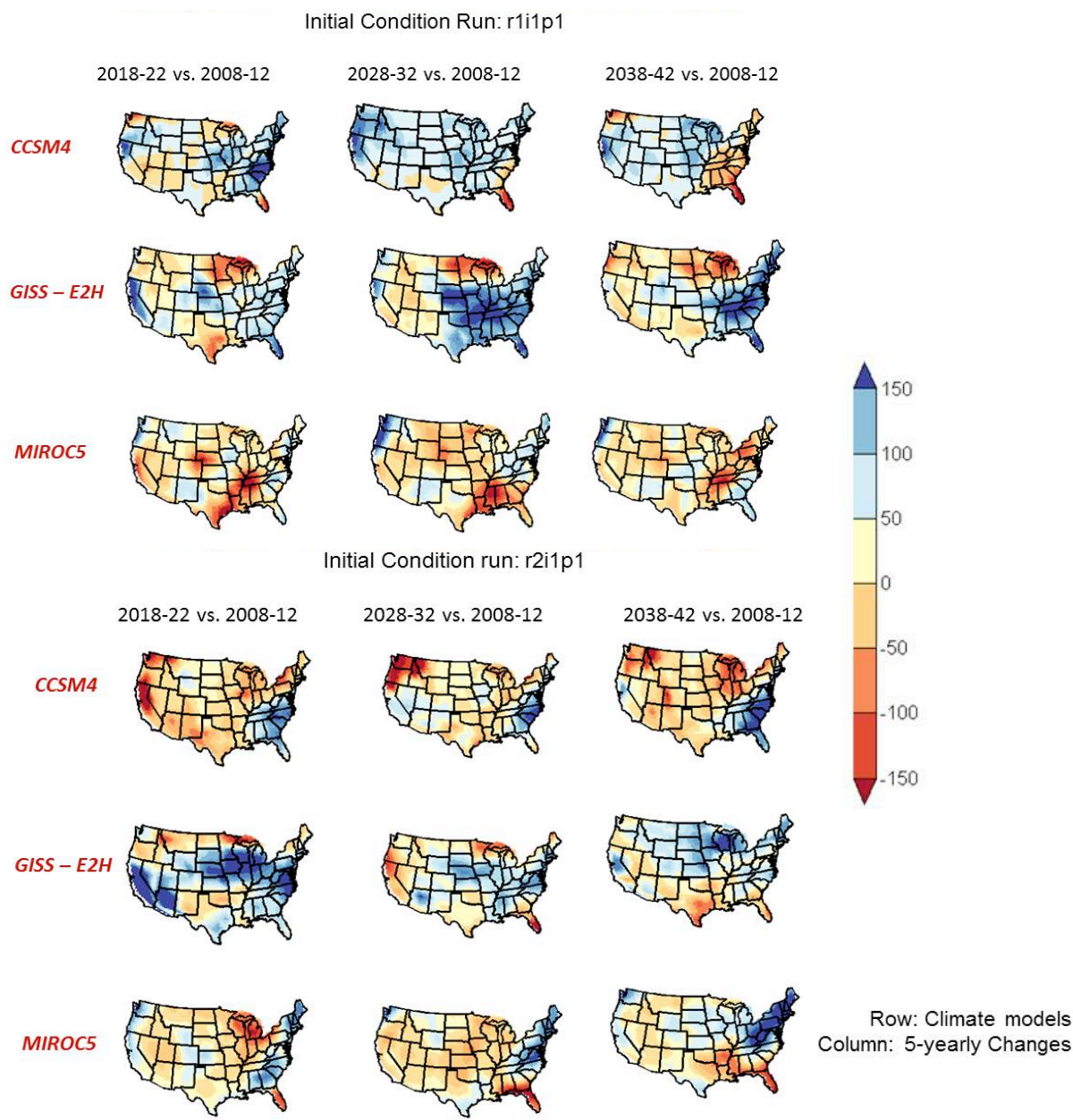

**Figure 3.3** Changes in freshwater availability across models and initial condition runs

Each of these combinations represents plausible future scenarios and all should be considered in the analysis. In this proof of concept work, we have just considered three climate models and two initial conditions; however, a comprehensive analysis should include all climate models, all initial conditions, and all GHG emissions scenarios that are available in CMIP5 archive.

***Uncertainty among Models and Initial Condition Runs in Estimate of WAACI***

Next we look at the estimate of WAACI index for MME minimum and MME median for 2030s and 2040s as shown in Figure 3.4. We can observe the intensification of water scarcity over many regions.

***Uncertainty in Runoff Estimate due to GHG Scenarios (RCP2.6 versus RCP8.5)***

Figure 3.5 shows estimates of freshwater availability for two GHG emissions scenarios (RCP2.6 and RCP8.5) for MME minimum in the 2010s (2008-2012). We observed drying patterns over the Midwest, Gulf coast and Southwest regions and wet patterns over the

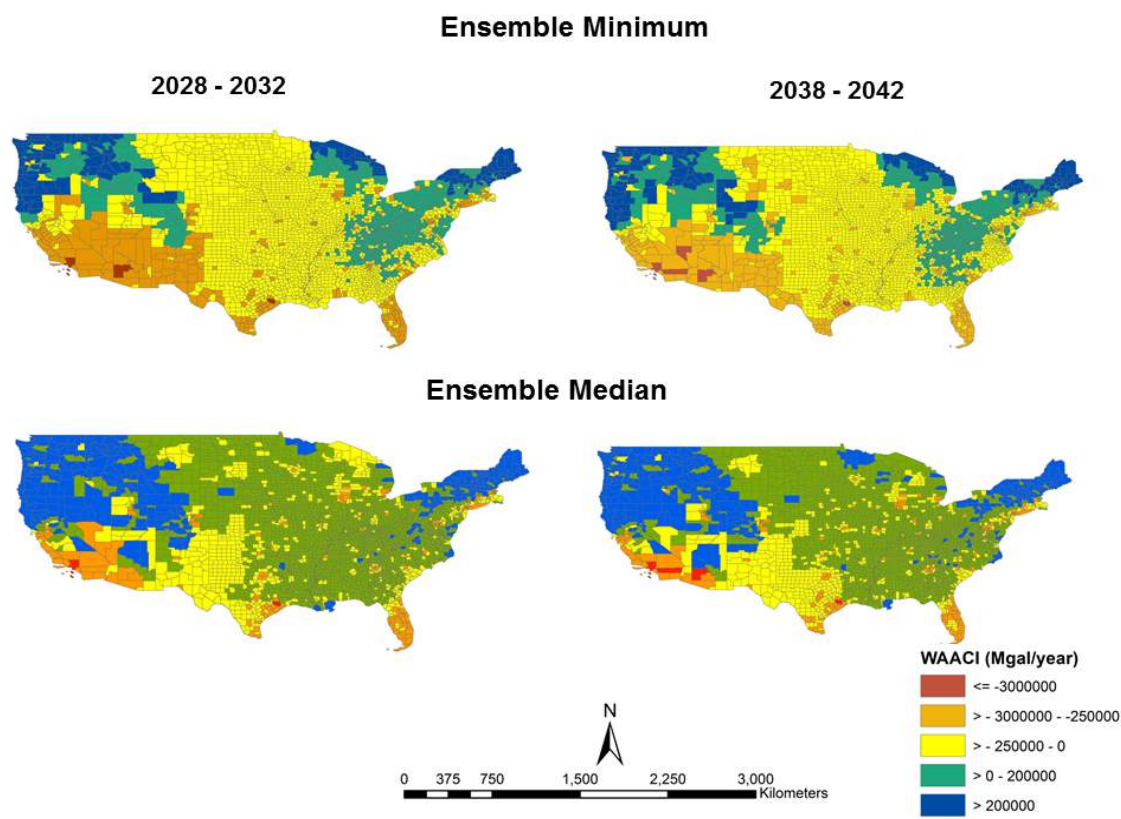

**Figure 3.4** WAACI during projected time windows computed using ensemble MME minimum and MME median

Northeast and coastal Northwest regions. Under RCP8.5 scenario, the area under drying pattern is large compared to RCP2.6. Figure 3.6 shows the spatial distribution of projected freshwater in the 2030s and 2040s for MME minimum. Figure 3.7 shows changes in $P - E$

Northeastern University

for the 2030s and 2040s with respect to 2010s. Drying patterns are observed over the Great Lakes and Central regions for RCP2.6 emission scenario as compared to wet patterns for RCP8.5.

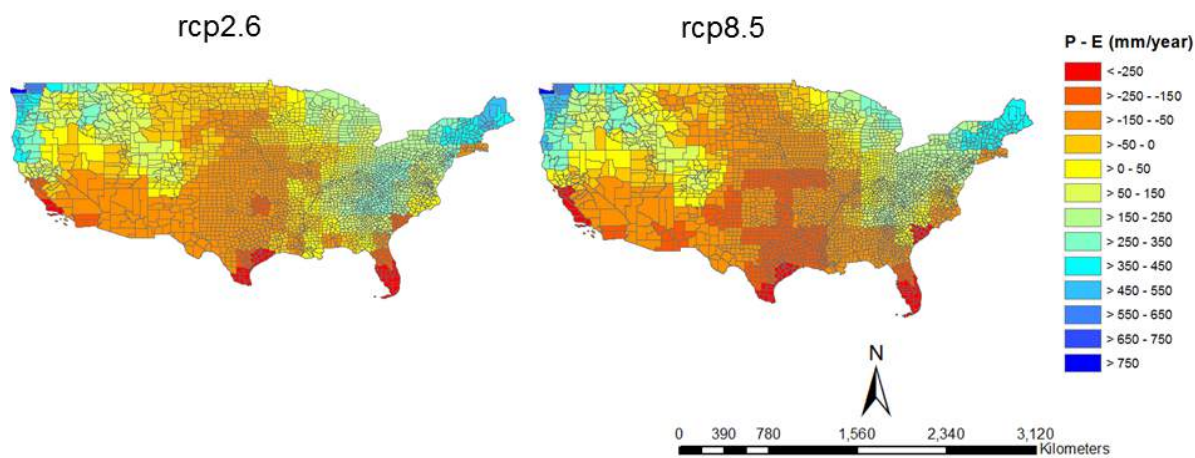

**Figure 3.5** Spatial distribution of runoff during 2010 in RCP2.6 and RCP8.5. *Note:* For RCP2.6, climate model GISS-E2H has only one initial condition run.

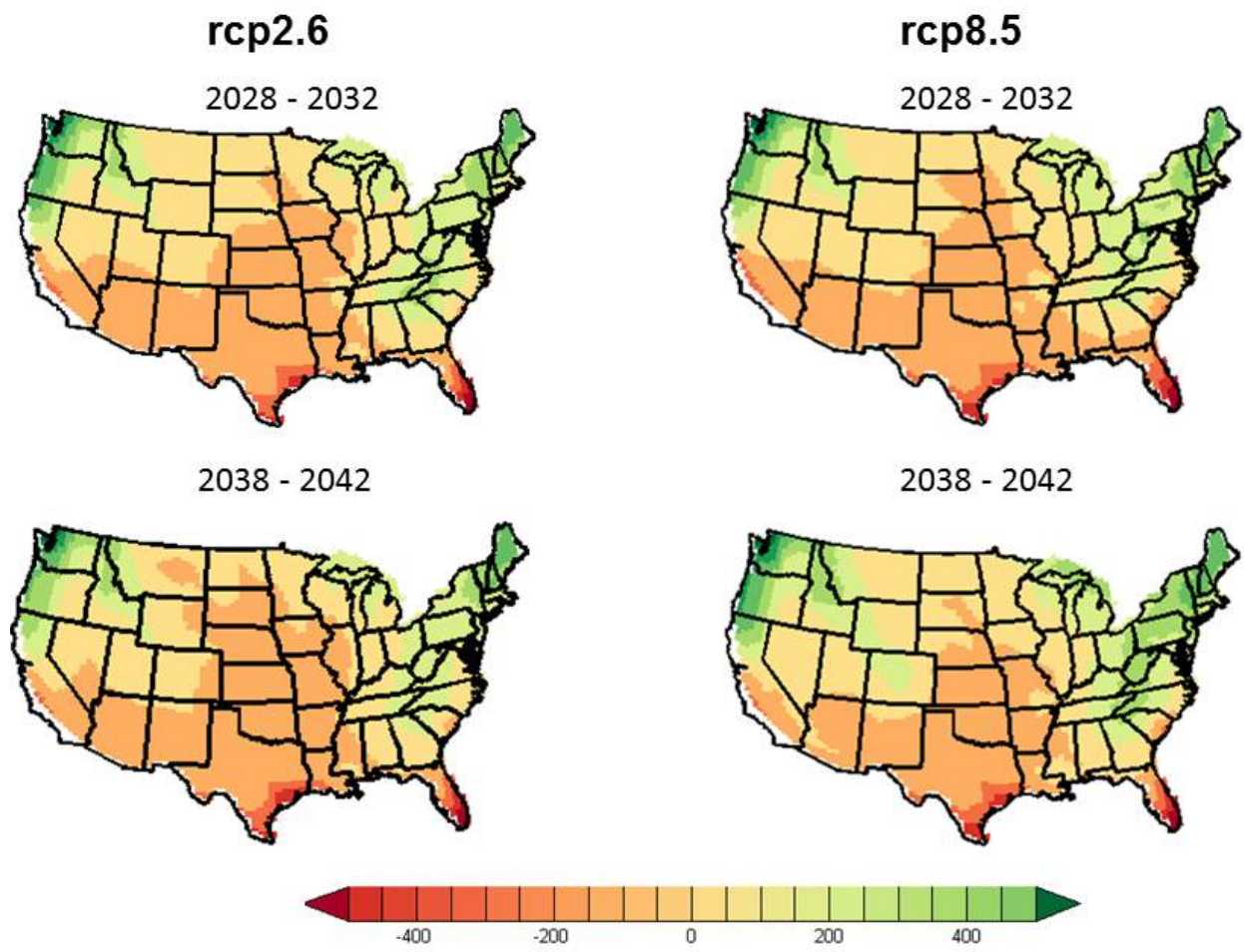

**Figure 3.6** Spatial distribution of runoff at 2030s and 2040s in RCP2.6 and RCP8.5

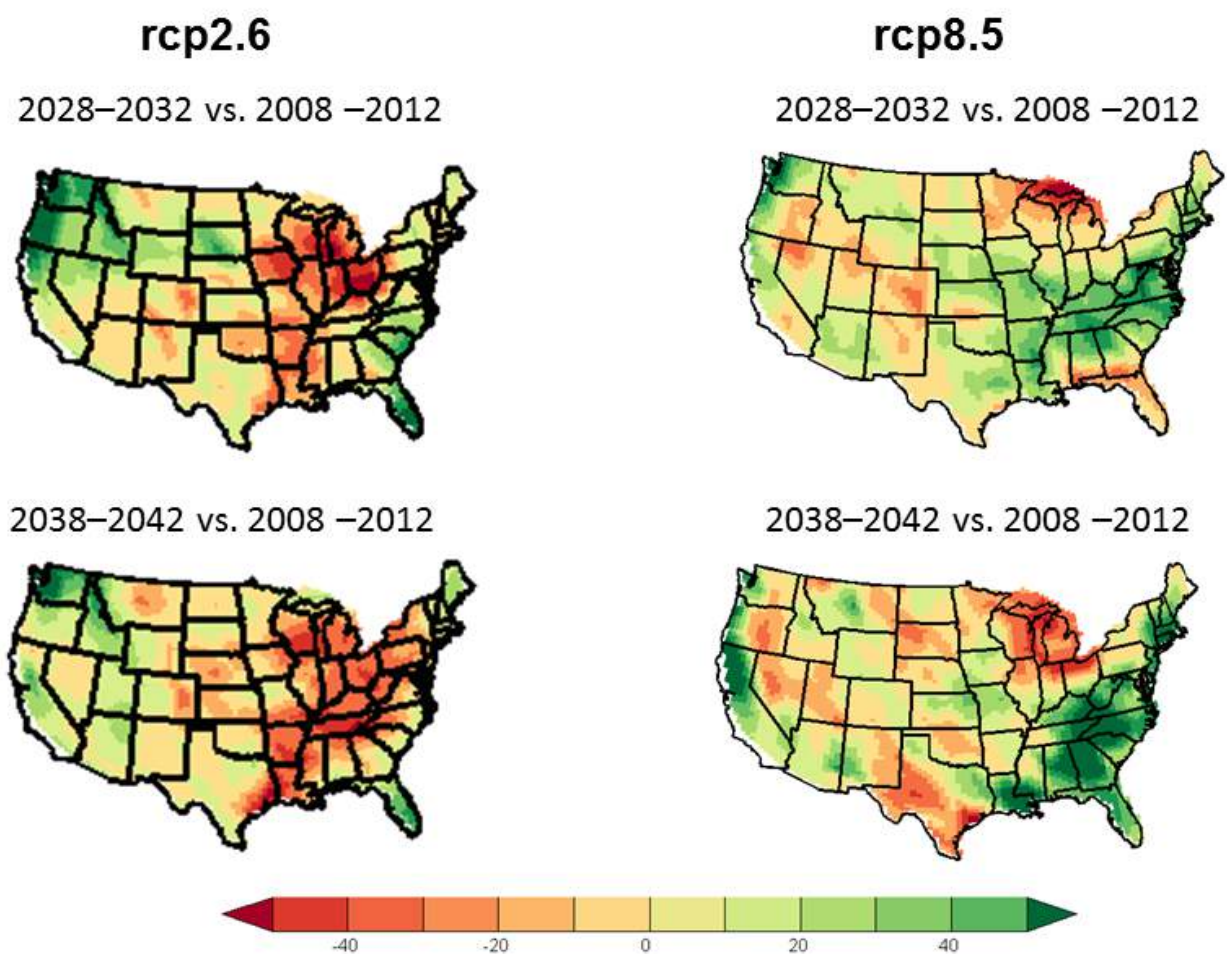

**Figure 3.7** Changes in spatial distribution of runoff at 2030s and 2040s in RCP2.6 and RCP8.5

**Uncertainty in WAACI during Projected Time Period**

A similar visual comparison is also performed for WAACI for two future time periods (2030s and 2040s) as shown in Figure 3.8. Drying patterns are observed in the Gulf coast, Southwest, and a few counties in Texas for RCP8.5. For example, the relative increase in exposure to water scarcity by 2040 under RCP8.5 is 1.31%. The analysis of changes in WAACI for both GHG emissions scenarios is shown in Figure 3.9. We observe substantial differences in drying (wet) patterns in some counties for both scenarios as shown in Figure 3.9.

Northeastern University

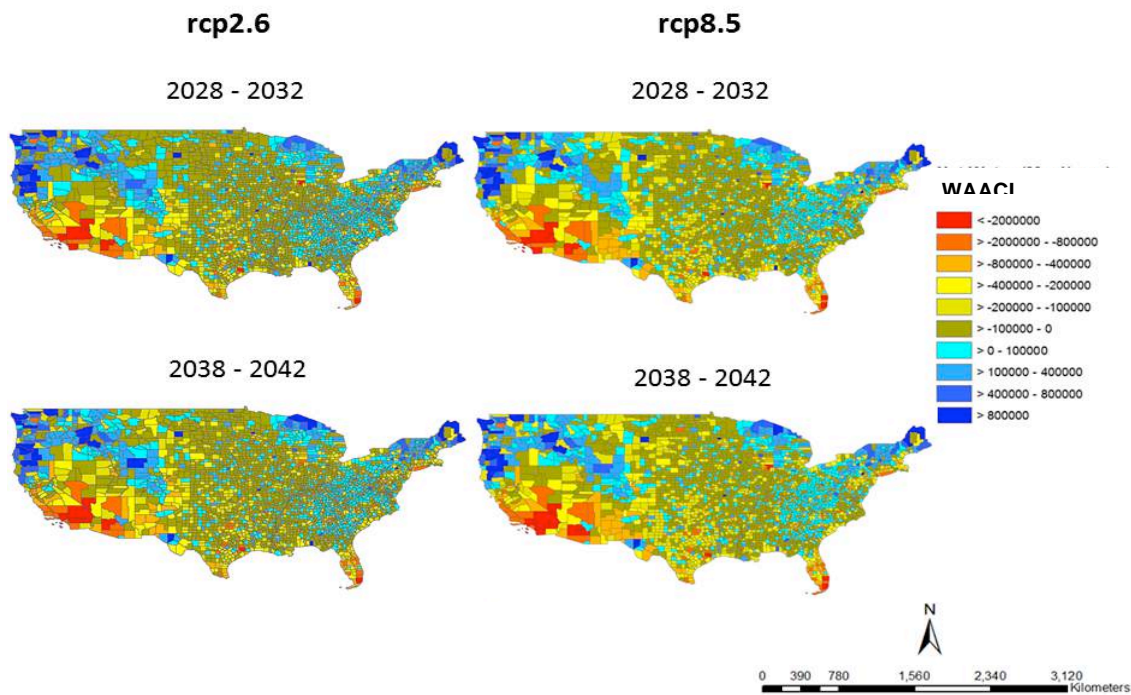

**Figure 3.8** Spatial distributions of projections of net water availability (fresh water available – demand) in multimodel minimum ensemble in RCP2.6 and RCP8.5

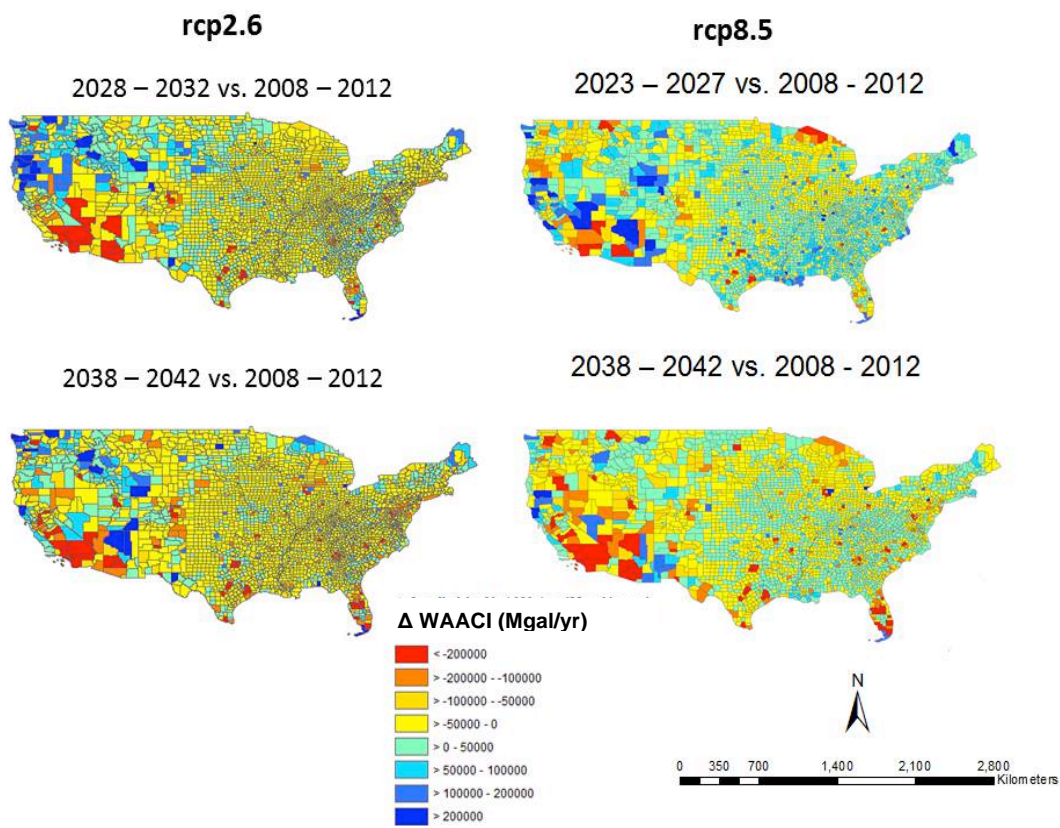

**Figure 3.9** Changes in net available water during projected time period in 2030s and 2040s in RCP2.6 and RCP8.5 emission scenarios

## 3.2 Increase in Stream Temperature

### 3.2.1 Trends in Historical Stream Temperature

Water temperature also impacts power generation; water at high temperature reduces the efficiency of power plant. We want to examine how the stream temperature will rise in the future under climate change. As mentioned above, we develop nonlinear regression models based on historical stream temperature and data from climate models to project water temperature. Before we look into the trends in projected stream temperature, we perform a non-parametric trend analysis (increase versus decrease in temperature) on historical stream temperature to gain some prior understanding about spatial patterns of water quality. We use the Mann-Kendall test with correction for ties and autocorrelation to perform the trend analysis.

Figure 3.10 shows the nature of trends (upward versus downward) at USGS stream gauge locations for historical stream temperature along with power plant capacity.

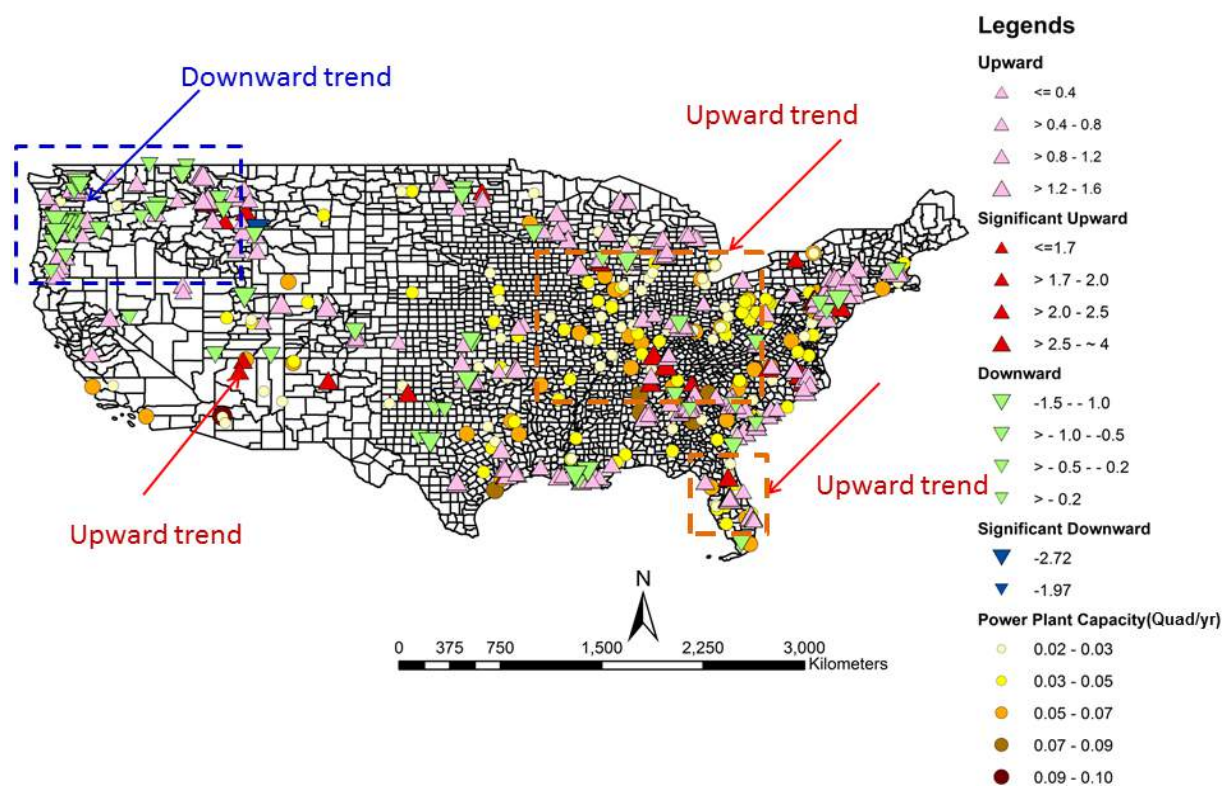

**Figure 3.10** Trends in stream temperature and higher capacity wet cooled power plants. The time period for analyzing trends may not overlap with each other for individual stations. Circles and triangles indicate spatial locations of power plants and stream gauges. The size (and shading) of the circle and triangle is proportional to the magnitude of power plant capacity and trend.

**Northeastern University**

SDS

Regions of increasing and decreasing trends with rectangular boxes are drawn in the Figure 3.10. At several locations a statistically significant increasing trend (at 10% significance level) is observed. A list of states and counties with increasing and decreasing trends is summarized in Table A5.

### 3.2.2 Trends in Projected Stream Temperature

Next we look at trends in projected stream temperature in the 2030s (2028-2032) and 2040s (2038-2042). Stream temperatures for future time periods are computed using Support Vector Regression as described in Chapter 2. Figure 3.11 shows spatial distribution of bias corrected maximum stream temperature in the 2030s and 2040s. Stream temperatures above 30°C are observed over Southern regions (Texas, Louisiana, Kansas, and Oklahoma), Gulf coast (Florida, Georgia, South Carolina, North Carolina) and in the Northwest (Oregon).

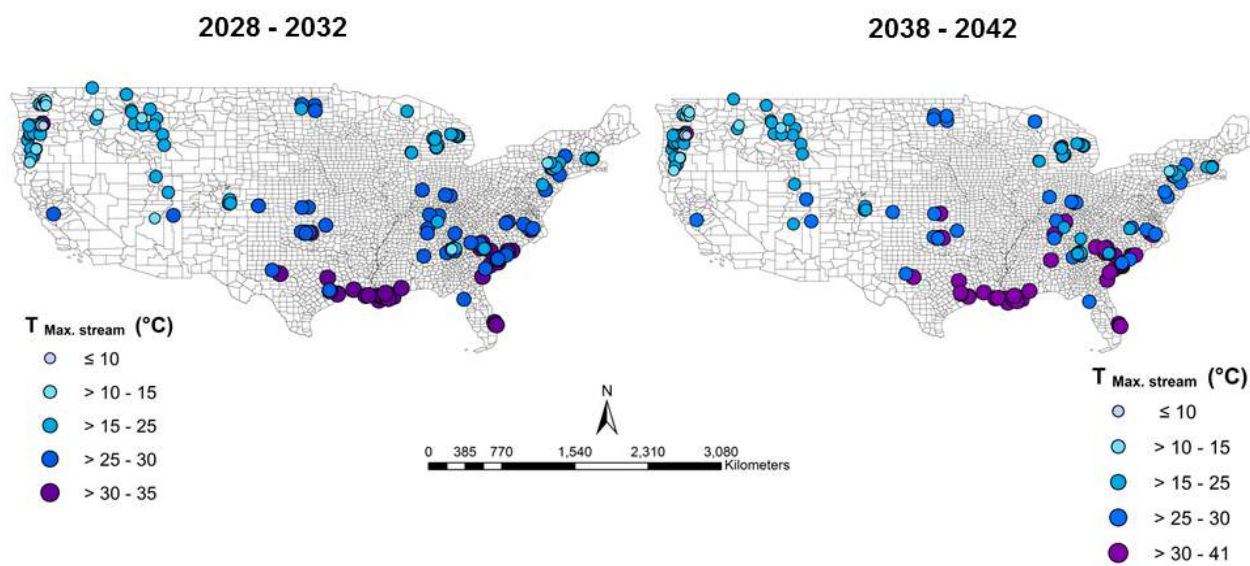

**Figure 3.11** Maximum stream temperature during projected time window

In addition, we also investigated changes in maximum stream temperature in the 2040s compared to the 2010s as shown in Figure 3.12(a). No uniform trend in stream temperature is observed. We also show changes in maximum stream temperature in the 2040s with air temperature in the background in Figure 3.12(b) to see whether any positive correlation exists between the air and stream temperature.

SDS

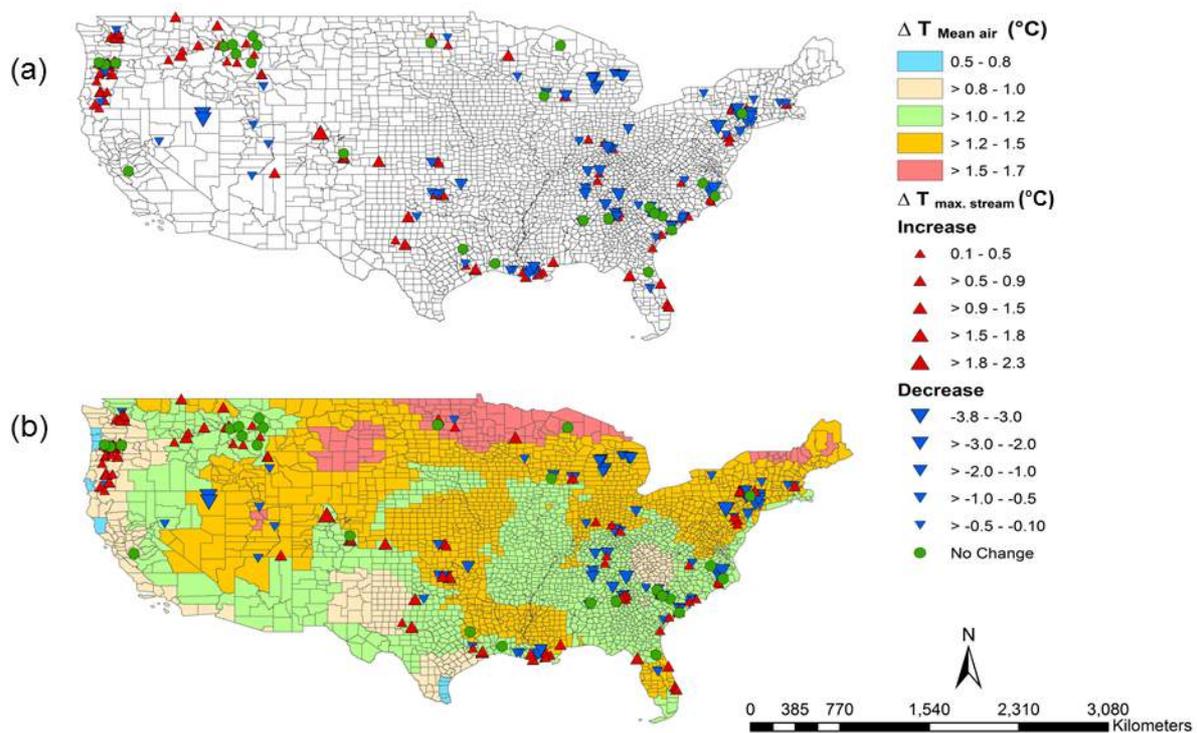

**Figure 3.12** Difference in changes in maximum stream temperature (2040 versus 2010) (*a*); same as in (*a*) but at the background of changes in mean air temperature (*b*)

# Chapter 4: Power Production at Risk

## 4.1 Power Production at Risk in 2030s and 2040s

We assessed the total power production at risk by aggregating annual production capacity of all power plants in the counties, where the WAACI index is negative and stream temperature is above the EPA prescribed threshold (Table A4). In 2030s, we assess power production at risk for two cases - MME minimum and MME median, and in 2040s only for MME minimum. We used MME median to build a regression relationship between average air and stream temperature. Maximum stream temperature is obtained from the projected monthly average stream temperature series. Figure 4.1 and Figure 4.2 show total power production at risk in the 2030s (2028-2032) and 2040s (2038-2042). As shown in Figure 4.1, power plants in the southwest U.S. and Florida are exposed to water scarcity (highlighted in red box). Few locations in the Southern U.S., such as South Carolina, Louisiana and Texas show evidence of water scarcity and warmer stream temperatures exceeding EPA limit.

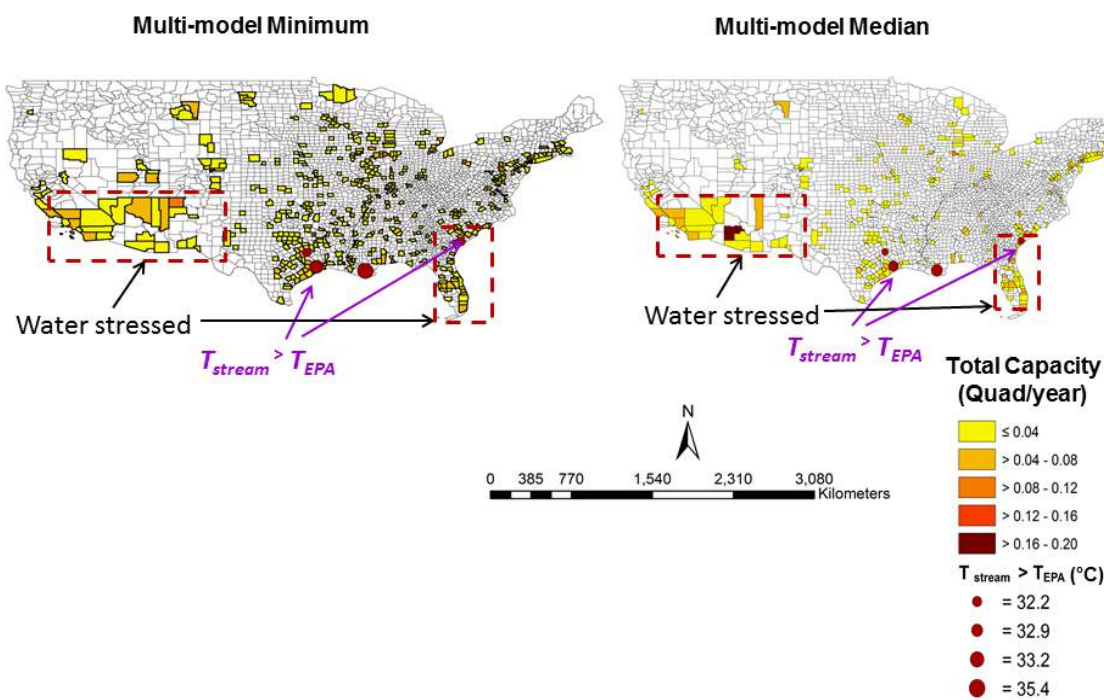

**Figure 4.1** Total power production risk due to decreasing water availability and increasing stream temperature for wet cooled thermoelectric plants during 2030

Northeastern University

SDS

In the 2040s, several stream gauge locations will most likely exceed the EPA prescribed threshold for water temperature as shown in Figure 4.3. However, a visual comparison shows that the combined effects of both water scarcity and high water temperature in the 2040s are similar to that in the 2030s. A list of counties that will most likely exceed the EPA regulations on water temperature is summarized in Table A6.

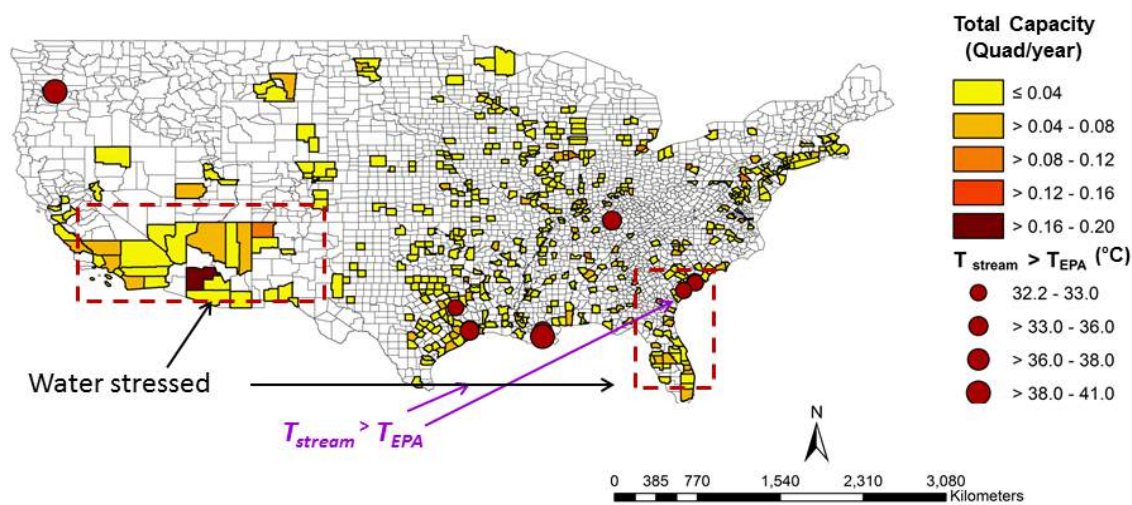

**Figure 4.2** The total power production risk due to decreasing water availability and increasing stream temperature for wet cooled thermoelectric plants during 2040. Water stressed region is identified using WAACI index, which is computed using ensemble minimum of climate models

Figure 4.3 shows yearly trends in total power production at risk for the wet cooled power plants for those counties where the WAACI index is projected to be negative in the 2030s and 2040s. Here we calculate water stress considering both MME median and MME minimum of climate model ensembles. The total capacity of power plants exposed to water stress exhibits an overall increasing trend over the decades.

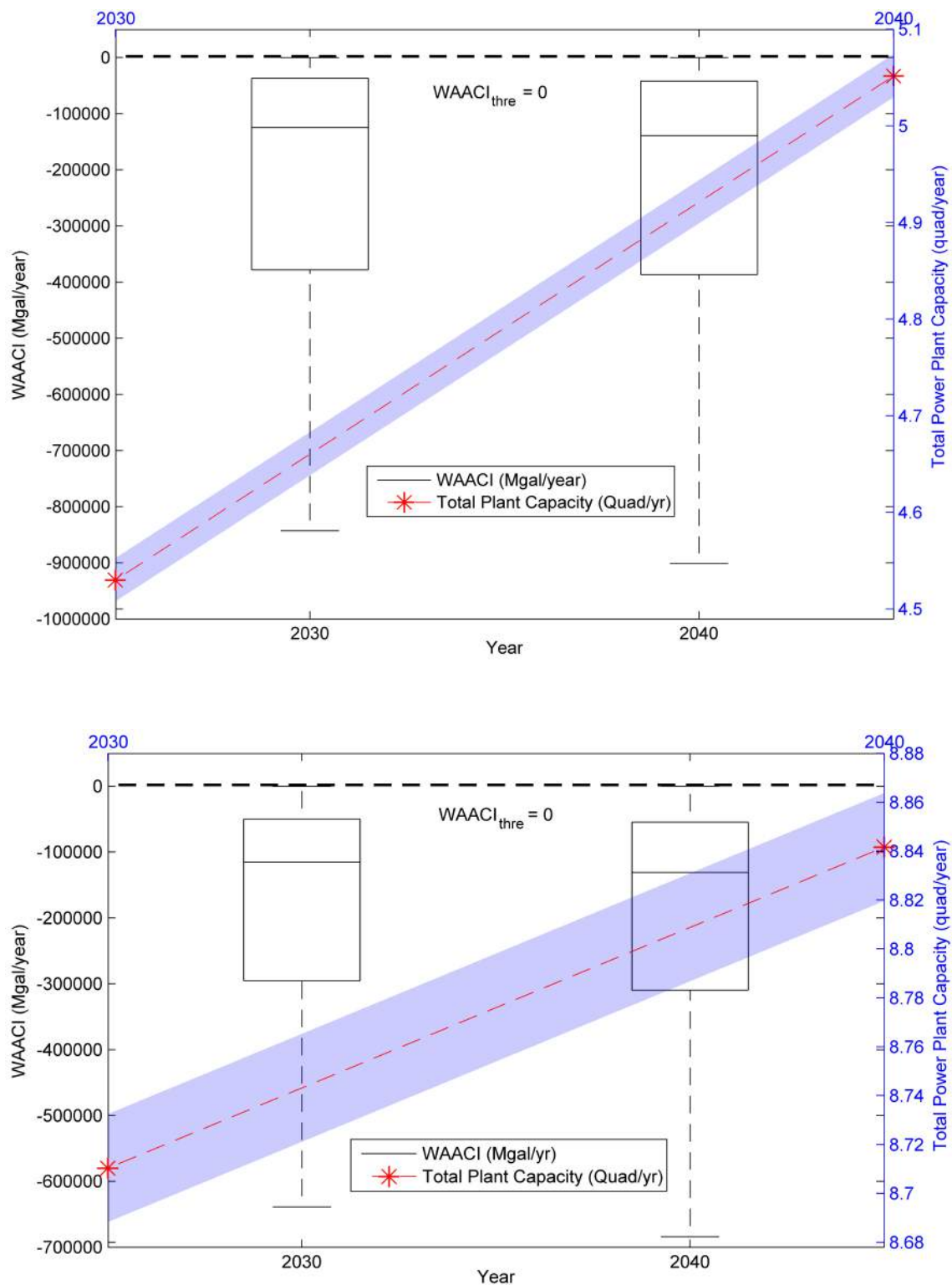

**Figure 4.3** WAACI versus total capacity of wet cooled thermoelectric plants in water stressed counties of Conterminous U.S. Shaded region shows ± 1 standard deviation (σ) of power plant capacity (top panel for MME median and bottom panel for MME minimum). Dotted line depicts threshold value to define "water stress"

Northeastern University

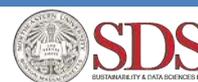

# Chapter 5: Conclusions and Discussion

Here we perform a preliminary investigation on the first order estimates of power production at risk from wet cooled thermal power plants due to water stress. The analysis is focused over the mainland U.S. at county scales for multiple 5-year time windows over the next 30 years. The total power production at risk is quantified by aggregating the total capacity of power plants at locations where the index defined for water availability (WAACI) is negative and the stream temperature is above a threshold prescribed by EPA. In this chapter, we briefly summarized the limitations of the proof of concept analysis, summarize and discuss the main results, and briefly discuss future work.

## 5.1 Limitations of Proof of Concept

o Future climate data to estimate freshwater availability (precipitation and evapotranspiration) are considered from only three climate models and two GHG emissions scenarios (RCP2.6 and RCP8.5); however, most of the analyses are performed with only RCP8.5.

o To consider the uncertainty due to climate internal variability, we considered climate data from only two initial conditions.

o Precipitation and evapotranspiration data from each climate models' native grid are bi-linearly interpolated to a common grid, 2-degree spatial resolutions, to estimate fresh water availability at regional scales. Water availability is estimated as the difference between precipitation and evapotranspiration at each grid point, and then spatially interpolated within ArcGIS to get estimates at county levels. Here we used a simple method of bilinear interpolation to estimate regional water availability; more robust estimates may be obtained by performing downscaled data (discussed below).

o Future water demand is considered only from municipal and domestic public supply; demands from other sectors are assumed as constant. Future water demand is estimated using projected population data and per capita water use. Future population is projected using 2010 U.S. Census population data and annual mean growth rates.

o For stream temperature projections, a nonlinear regression approach is developed using downscaled air temperature data as the predictor. The effects of exogenous

Northeastern University

SDS

factors such as anthropogenic changes or effluent discharge are not considered.

o  Power production at risk is estimated by aggregating the installed capacity of power production for all those counties for which the water stress index is negative and water temperature exceeds the threshold specified by the EPA. A visual risk analysis is performed combining water scarcity and stream temperatures trends with the capacity of power plants in the spatial proximity.

o  We have not validated our results with hydrological models; whether to estimate water stress (such as application of water balance hydrological models) or to estimate stream temperature (application of numerical model based on heat transport equation).

o  We present a first order estimation of water stress approximating surface water availability/runoff as the difference between precipitation and evaporation (from underlying surface and vegetation). However, withdrawal of alternate water sources such as groundwater, desalinated seawater in thermoelectric cooling is not considered.

## 5.2 Summary of Results for the Proof of Concept

The main findings from this study are summarized below:

o  First, we visually compared spatial patterns of estimates of freshwater from three climate models against the reference data, ERA-Interim, in the 2010s. We observe noticeable differences in estimates of freshwater across several regions between the climate model and reanalysis data and in some cases among the models themselves. The inter-model differences over some regions are complete opposite in sign; i.e., one model showed drying patterns over some regions while another showed wet patterns over the same region (Figure 3.1). These contrasting and opposite results make a case for performing the analysis using data from multiple climate models, possibly from all climate models available in the CMIP5 archive.

o  We defined a metric based on water supply and water demand (only municipal public and domestic supply are considered), WAACI, to quantify water stress. We computed this metric for MME minimum and MME median at county scales to get average and worst-case scenario estimates of water stress. Most regions in the Midwest and Southwest are under water stress while some regions in the Pacific Northwest and Northeast have surplus water (Figure 3.2).

o  We examined the uncertainty in the estimates of fresh water from multiple

combinations of models, initial conditions, and GHG emissions scenarios in the 2020s (2018-2022), 2030s (2028-2032) and 2040s (2038-2042) as shown in Figures 3.3-3.9. The uncertainty analysis is mostly limited to qualitative evaluations by visual comparison of spatial maps. Although a wide range of uncertainties is observed with even a small number of models, initial conditions, and GHG scenario combinations, we need to further explore the uncertainties with all plausible combinations of models, initial conditions, and emissions scenarios.

o   We developed nonlinear regression models to project future stream temperatures. An upward trend in stream temperature is observed over most regions. The difference between maximum stream temperature during the projected and current period show decrease in stream temperature over many locations in the Northeast and an increase in the Northwest and Gulf coast parts of the US.

o   Our analyses suggest that in the near term, more than 200 counties in the contiguous United States are likely to be exposed to water scarcity for the next three decades. Further, we noticed that stream gauges in more than five counties in the 2030s and ten counties in the 2040s showed a significant increase in water temperature, which exceeded the prescribed EPA limits. By superimposing the location of power plants with capacity over spatial maps of water scarcity and water temperature, we found that the power plants in South Carolina, Louisiana, and Texas are most likely vulnerable owing to climate driven stresses.

**5.3 Discussions of Proof of Concept**

Thermoelectric power plants generate approximately 91% of total power production in the United States and accounts for about 40% of water withdrawals (Cooperman et al., 2012), which is highest among all sectors. Cooling water used for steam condensation shares about ~ 98% of total power plant water use (EIA, 2013), and varies based on fuel source, power generation technologies, cooling system used, and other climatic and external factors (Macknick et al., 2011; Stillwell et al., 2011). Previous studies (King and Webber, 2008; Roy et al., 2012; Sovacool and Sovacool, 2009) have shown that power production from thermoelectric plants in many regions of the U.S. are at risks primarily due to decreasing freshwater availability driven by climate change and multi-sector demands. Most of these studies were based on data from an older generation of climate models and associated

Northeastern University

SDS

greenhouse gas emissions scenarios and did not consider climate internal variability. In addition, these studies were focused on continental-to-global scales and from mid-to-end of the century timescales. In addition, the combined effects of low flow and increasing trends in stream temperature on power production were not studied before. A decline in summer stream flow leads to a further rise in stream temperature, because with low flow reduces the thermal mass of water, causing streams to heat up more quickly than under higher flows.

The impact of climate change and population growth on the water-energy nexus problem has been studied before. A fundamental difference here is the consideration of decadal (30-year) time horizon at 5-yearly increments. This contrasts with the mid- to end- century scale projections in most prior work (Blanc et al. 2014; Hejazi et al. 2014; Roy et al. 2010). The combination of two factors, specifically (a) predominance of what may be viewed as "deep uncertainties" in climate relative to the warming trends, and (b) the relatively near-term needs of the stakeholders, makes our problem challenging and urgent. We further note that while 30 years is often considered a time-scale suitable for assessment of average climatology, the typical length of planning horizons is also 30 years for multiple stakeholders.

Per ARPA-E recommendation, we have computed freshwater availability from GCMs at county levels for multiple 5-year time windows for the next 30 years (2010-2040). At these decadal to multi-decadal time horizons, internal variability (uncertainty due to small difference in initial conditions of the earth systems) dominates over model uncertainty (uncertainty due to lack of understanding of physics) and scenario uncertainty (uncertainty due to the insufficient knowledge of amount of future greenhouse emissions). This requires us to consider climate variables not only from multiple models but also from an ensemble of initial conditions. We are not aware of any studies of the water-energy nexus problem in which climate data from multiple initial conditions have been considered to assess power production at risk. We are aware of only one study in which the effect of water temperature on power production was considered (van Vliet et al., 2012). In this study, they calculated changes in water temperature using hydrologic analysis based on a deterministic model for 30 years periods at the end of this century. Here we developed a nonlinear regression model based on downscaled climate variables to project changes in stream temperature over next 30 years. However, in this preliminary investigation, we have not explored data from combinations of all climate models and initial conditions.

Northeastern University

Obtaining credible estimates of water supply and water temperature at the scale of power plants is another major challenge. The data from CMIP5 climate models are available at spatial resolutions on the order of 110 km - much larger than the size of a reservoir. Statistical downscaling is performed to downscale climate data from coarser resolutions to local scales to perform impact assessments. We have not performed statistical downscaling in this work to estimate fresh water availability but we will explore this option in future work; however, the tradeoff with downscaling is that it adds additional uncertainty. Besides temperature and precipitation, statistical downscaling of other meteorological variables is little known in existing literature, so we have to be cautious in downscaling evapotranspiration as it is used in the estimation of water availability. Reliable estimates of water supply at regional scales from climate models can still be made to get an understanding of power production risk at an aggregate level; however, projection of stream temperature needs to be done at local scales (preferably close to the location of the reservoir from where water is withdrawn). Presently we have developed nonlinear regression-based models to project stream temperature using mean air temperature as a predictor; we can improve on the selection of predictors and can extend the list of predictors. The estimation of future water demand from multiple sectors is another challenge. Presently we have considered future water demand from only municipal domestic and domestic supply. In future work we will project demands from other sectors as well. Despite these technical challenges we have demonstrated a proof of concept on how risk assessment can be performed to quantify power production at risk over the next 30 years. We propose recommendations to build upon the existing analysis for a robust assessment of power production at risk. However, we presented our results under a few strong assumptions as discussed above. This proof of concept demonstrates how to handle these challenges in a comprehensive manner. The study should be further expanded to a comprehensive and consistent scenario analysis, including changes in extremes, and their potential consequences on power plants' water demand and associated cost-benefit tradeoffs.

## 5.4 Future Work

There are many ways in which the existing analysis can be extended to assess power production from existing power plants that is at risk due to water stress. We suggest both short-term and long-term solution strategies. We can perform an immediate analysis by using

Northeastern University

SDS

the methods and tools used in this proof of concept work but by considering an extended set of future climate data from multiple models (more than 30 GCMs in the CMIP5 archive), all GHG emissions scenarios (4 RCPs), and initial conditions (more than 300 for all GCMs and 4 RCPs). In the extended analysis, we will estimate future water demand for other sectors as well. We are working on developing a statistically grounded methodology to delineate different sources of uncertainty so that the effects of model uncertainty, scenario uncertainty, and internal variability on power production can be studied separately. We have discussed the solution strategies in details in the next chapter on challenges and the way forward.

# Chapter 6: Challenges and the Way Forward

## 6.1 Challenges

Our analysis showed that electricity generation from thermoelectric power plants in certain regions of the United States are at risk due to diminished freshwater availability and increased stream temperature. Considering that future electricity demand will grow, factors such as climate change and population growth most likely will exacerbate the problems in the regions already affected due to water stress. In the U.S. in 2007-2008 during the warm and dry summers, several power plants had to reduce their installed electricity generation capacity and shut down for several days due to insufficient water available for cooling and environmental regulations on hot water discharges (NREL, 2011; NETL 2009). There are concerns that these events might become more frequent, intense, and widespread in the future. The latest generation of climate models and observation data sets show an unequivocal increase in average temperature (IPCC, 2013), which has a direct impact on available water resources. Here we highlight the major challenges that should be addressed in the follow up work to perform a robust assessment of the vulnerability of energy sectors in the future due to climate change and variability.

o   **Spatial Resolutions:** Data from the current generation of climate models are not credible to perform an impact assessment at spatial scales relevant for power plant operations. To overcome this problem, downscaling of atmospheric variables from GCMs is performed to obtain weather and climate at regional or local scales. Two downscaling approaches - statistical[5] and dynamical[6] – have been discussed extensively in the literature; downscaling of climate models' output at regional scales introduces additional uncertainty. Owing to its physical consistency and interpretability, dynamical downscaling is often preferred to statistical downscaling. However, the former is computationally expensive (it usually takes many days, sometimes months, to run an analysis even over a small region in the U.S.) and often does not capture climatic teleconnections (Boulard et al., 2013; Hudson and Jones,

---

[5]   In statistical downscaling, a statistical relationship is developed from observations between large-scale variables and regional/local variables under historical conditions; subsequently, data from climate models are used as input to get the corresponding regional/local climate.
[6]   In dynamical downscaling, a regional climate model (RCM) in higher spatial resolution is run, which in turn is able to simulate local conditions more realistically. RCM is constrained by a global circulation model (GCM) at the boundary of the region of interest. Dynamical downscaling is computationally more expensive than Statistical downscaling.

Northeastern University                                    SDS

2002). In addition, recent studies even have questioned the value-addition of regional climate models (RCMs are employed to perform dynamical downscaling) over global circulation models (GCMs) (Kerr, 2013; Racherla et al., 2012). On other hand, statistical downscaling techniques are relatively computationally inexpensive and are based on statistical relationships between closely related atmospheric variables. However, statistical downscaling may not necessarily guarantee physical consistency between the variables and the relationships may change in a nonstationary climate (Milly et al., 2008). In future work, we will use the statistical downscaling technique with recent methodological advancements to estimate regional water availability and stream temperature.

o **Multi-sectors Water Demand:** In the present work, we have considered future water demand from only one sector (municipal public and domestic water), and we assumed that water demand from other sectors (irrigation for agriculture, industrial, mining, livestock, and aquaculture) will not change. In the future work, we will consider water demands from multiple sectors.

o **Internal variability:** Uncertainties in projected climate variables primarily result from three sources: model uncertainty, scenario uncertainty, and climate internal variability. Model uncertainty arises from the lack of our understanding of physics and imperfect numerical modeling of atmospheric processes. Insufficient knowledge on how much greenhouse gases will be emitted in the future gives rise to scenario uncertainty. The sensitive dependence on initial conditions of the earth system's variables combined with nonlinear-coupled interactions among different components of earth system gives rise to climate internal variability. The different sources of uncertainty dominate at different time scales; however, specifically at decadal to multi-decadal scales (0-30 years), internal variability and model uncertainty dominate over scenario uncertainty (Stocker et al. 2013). The dominance of internal variability at regional to local spatial scales is even more pronounced. Here we illustrate their importance and role through some examples taken from the IPCC fifth assessment report (AR5) from working group I (Kirtman et al. 2013). Figure 6.1 illustrates the relative importance of different sources of uncertainty for the projected global mean temperature; it can be observed that internal variability dominates over model and scenario uncertainties until the 2040s. Figure 6.2 illustrates the fraction of total variance explained for seasonal (December-January-February and June-July-August) and decadal mean temperature and precipitation at multiple spatial scales (Global, Europe, and East Asia). Some of the key points are: (1) the uncertainty in near-term projections is dominated by internal variability and model uncertainty, (2) internal variability becomes increasingly important on smaller spatial and time scales, (3) for projections of precipitation, scenario uncertainty is less important and internal variability is generally more important than that for air temperature. In the follow up work, we will include data from all initial conditions from the CMIP5 archive, as each of them represents a possible future climate state.

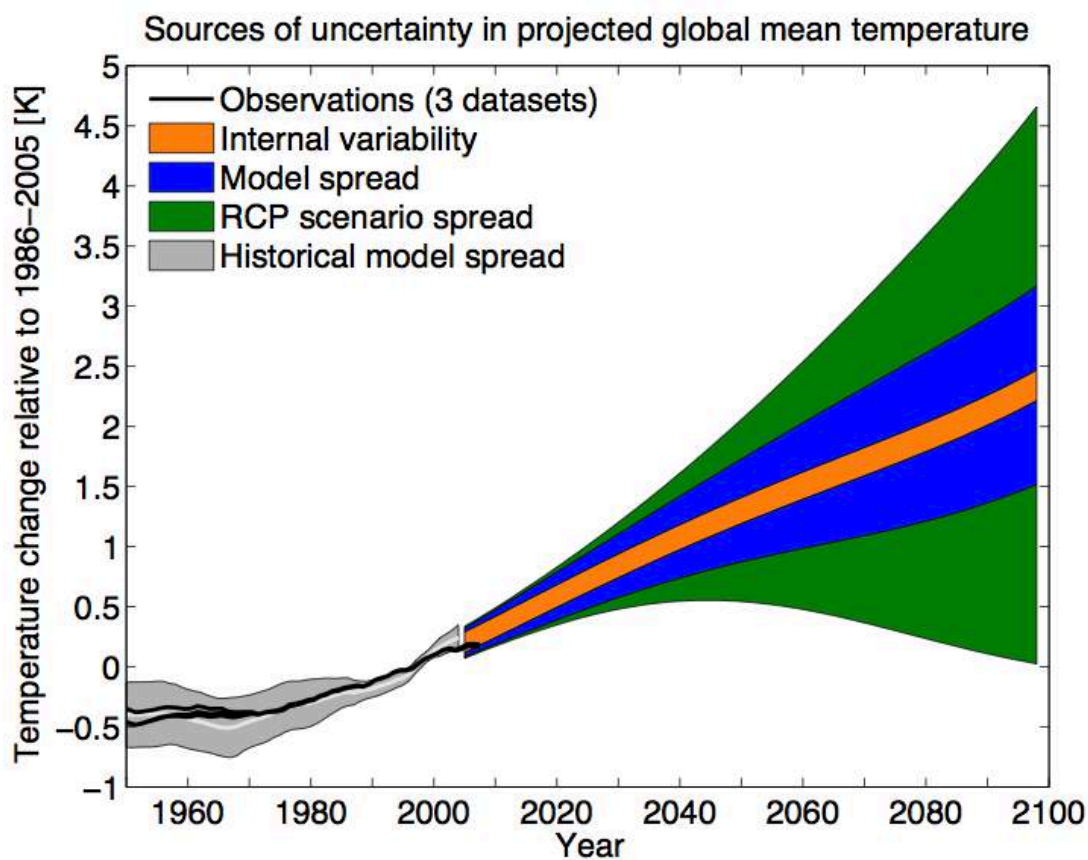

**Figure 6.1** Sources of uncertainty in climate projections as a function of lead time based on an analysis of CMIP5 results. Projections of global mean decadal surface air temperature to 2100 together with a quantification of the uncertainty arising from internal variability (orange), model spread (blue) and RCP scenario spread (green). (Source: IPCC AR5 Working Group I report, Figure 11.8, Page no 97)

o  **Model Uncertainty:** We discussed about model uncertainty in the previous section and illustrated their importance through Figures 6.1 and 6.2. Here we show examples of model uncertainty including internal variability through our earlier work on freshwater availability. Figure 6.3 illustrates differences in spatial patterns of *P-E* across 6 climate models including MME mean (top panel) and of precipitation and temperature across three initial conditions for one model, CCSM4 (bottom panel).

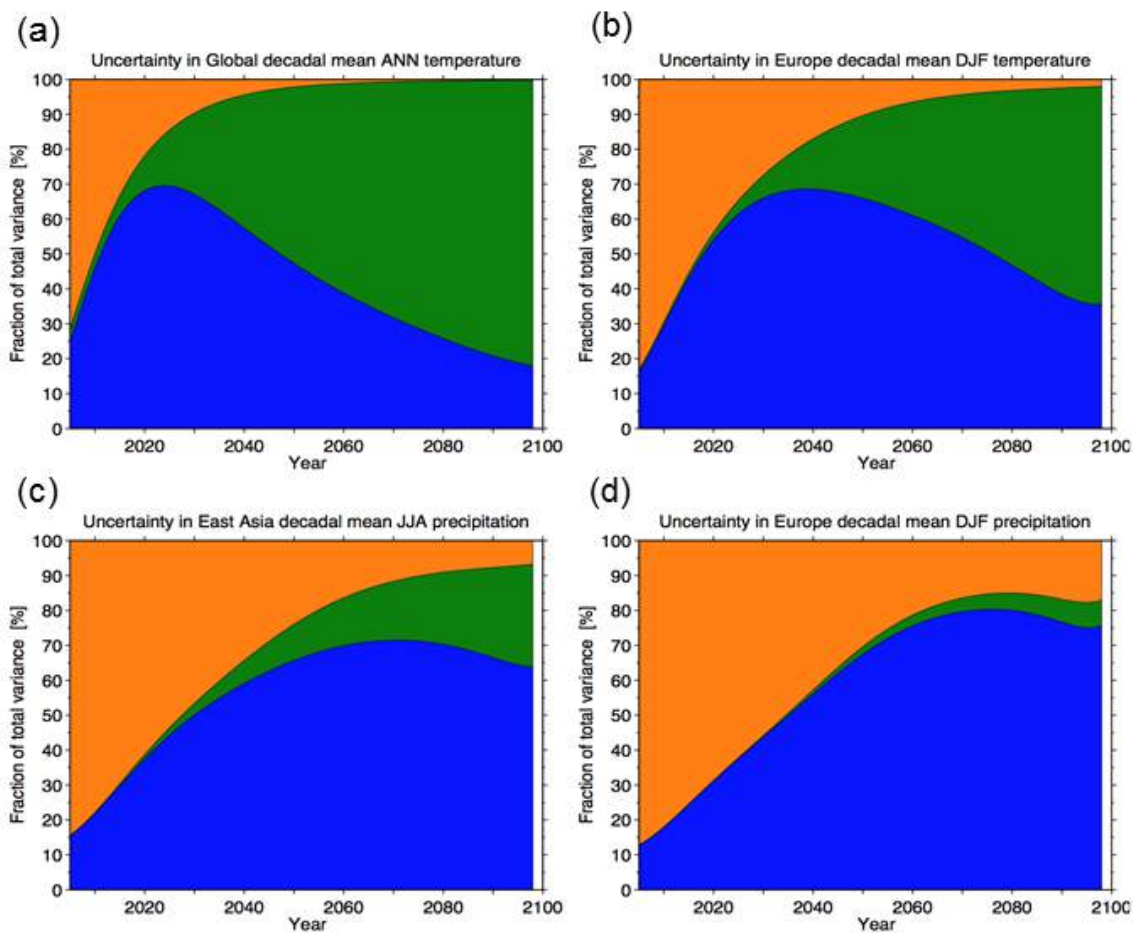

**Figure 6.2** The fraction of variance explained by each source of uncertainty for (a) global mean decadal and annual mean temperature, (b) European (30°N to 75°N, 10°W to 40°E) decadal mean boreal winter (December to February) temperature, (c) East Asian (5°N to 45°N, 67.5°E to 130°E) decadal mean (June to August) precipitation and (d) European decadal mean winter precipitation. (Source: IPCC AR5 Working Group I report, Figure 11.8, Page no 979).

In the top panel of Figure 6.3, we observe that CCSM4, FGOALS-S2, MIROC-ESM, and MME mean show dry patterns in P-E over the Pacific Northwest while GISS-E2 shows drying patterns. Similarly FGOALS-S2 shows drying patterns over the Northeast region, while other models including MME show wet patterns. In the top row of the bottom panel of Figure 6.3 (for three initial conditions from the same model), we see opposite spatial patterns of precipitation over several regions. These examples highlight the importance of including data from multiple climate models.(Tebaldi and Knutti, 2007). In the proposed work, we will consider data from all climate models and all initial conditions.

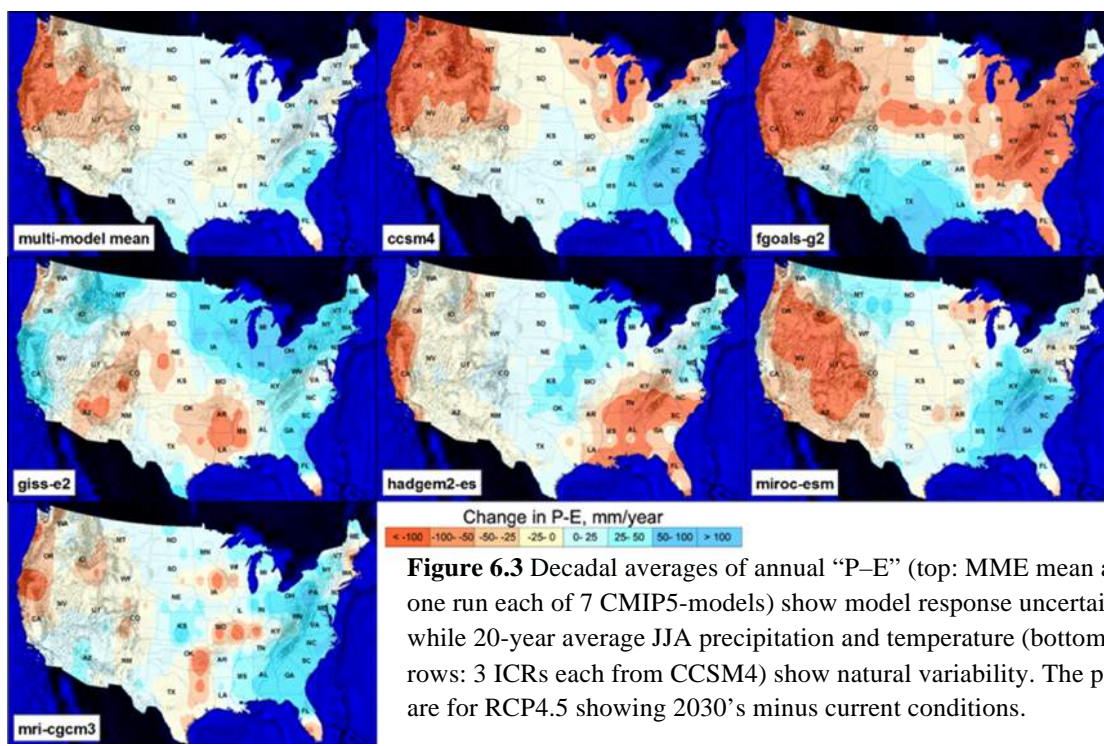

**Figure 6.3** Decadal averages of annual "P–E" (top: MME mean and one run each of 7 CMIP5-models) show model response uncertainty while 20-year average JJA precipitation and temperature (bottom two rows: 3 ICRs each from CCSM4) show natural variability. The plots are for RCP4.5 showing 2030's minus current conditions.

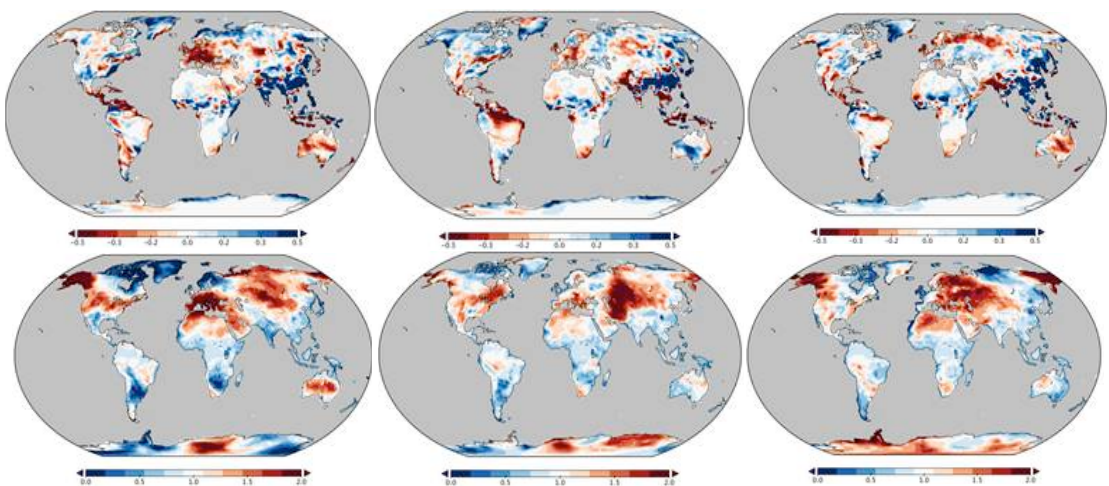

*Ganguly et al. 2013 (Unpublished)*

o **Cost-benefit tradeoffs:** A possibly greater source of uncertainty may arise from the need to incorporate cost factors that may be both time-phased (e.g. time value of money, as well as societal costs of environmental degradation and hazards to habitations) and geographically disparate. While the projections of future water stress together with an assessment of their uncertainties is relevant in science perspectives, the costs of reduced resilience of socio-ecological systems may be harder to quantify and occasionally have to rely on a set of relatively subjective metrics. The conventional approaches to probabilistic risk assessment usually begin with a null hypotheses test of "no trend" and its likelihood, with little or no attention given to the likelihood that we might ignore a trend if it really existed. Inference from such statistical tests can result in both Type I and Type II errors. The traditional decision-making process for adaptation decisions just considers Type I errors in cost-benefit analyses. The societal consequences of making such a mistake are over-investment in terms of unnecessary use of resources in building and maintaining infrastructure that is not required. In contrast, ignoring Type II errors in the decision-making process causes society to be underprepared for catastrophic losses. The risk based decision management involves tradeoffs between over-investment and under-preparedness at installation-level to develop short- and longer-term adaptation strategies: inform future infrastructure standards, design requirements, and guidelines for the reliable operation of power plants.

## 6.2 The Way Forward

We propose several steps to build on the existing work to extract robust actionable insights by considering additional plausible future scenarios and by using an improved methodology to compute metrics related to water supply and stream temperature. We have broken down the proposed recommendations into several tasks as summarized below.

### 6.2.1 Statistical Downscaling

Estimates of regional water availability and stream temperature can be improved by the use of statistically downscaled data. The performance of statistical downscaling depends on covariate relations, in which more reliably projected variables are used as predictors to enhance the projections of underlying climate variables. The set of predictors may include local and regional atmospheric variables as well as global climate indicators and oscillators. A combination of classification and regression approaches can be augmented with a mixture of experts to assign probabilities of individual predictands in the regression relationship (Das et al., 2014). In addition, our prior experience examining data from novel methods for enhanced statistical downscaling, based on physics-guided data mining techniques (Chatterjee et al., 2012; Das et al., 2014; Ganguly et al., 2014; Steinhaeuser et al., 2012, 2011) may prove useful in this direction.

### 6.2.2 Uncertainty Characterizations

We will employ climate data from multiple models, GHG emissions scenarios, and initial conditions from the CMIP5 archive to characterize the full range of uncertainties in the projection of freshwater supply and stream temperature, and subsequently on the risk to power production.

a. **Model Response:** Explore the range of behavior arising from our lack of understanding of the physics or model parameterizations, as captured through multimodel ensembles (MMEs):

1. Blend historical model skills with MME consensus through Bayesian (e.g., Smith et al., 2009), empirical (Santer et al., 2009), and process-oriented (Overland et al., 2011) methods.

2. Incorporate physical constraints (e.g., Fasullo and Trenberth, 2012; Sugiyama et al., 2010) and multivariate correlation structures (e.g., Liu et al., 2012; Tebaldi and Sansó, 2009), as well as novel data mining methods (Chatterjee et al., 2012; Steinhaeuser et al., 2012, 2011).

b. **Natural Variability:** Characterize the (nonlinear) dynamical behavior of the earth system, as determined from multiple initial condition runs (ICRs) (Deser et al., 2014, 2012) across model runs:

1. Bounds on predictability (e.g., Franzke, 2012; Shukla, 1998), for multiple variables, space-time aggregations and different seasons or regions, based on chaos and information theoretic methods applied to integrated earth system models (IESM) outputs (Branstator et al., 2012; Branstator and Teng, 2010).

2. Methods to characterize nonlinear dependence structures, including long-memory and long-range processes, through developments in nonlinear dynamics such as mutual information based associations (e.g., Reshef et al., 2011) and complex networks (e.g., Donges et al., 2009).

A flowchart of summarizing the details of this task is shown in Figure 6.4.

Northeastern University

SDS

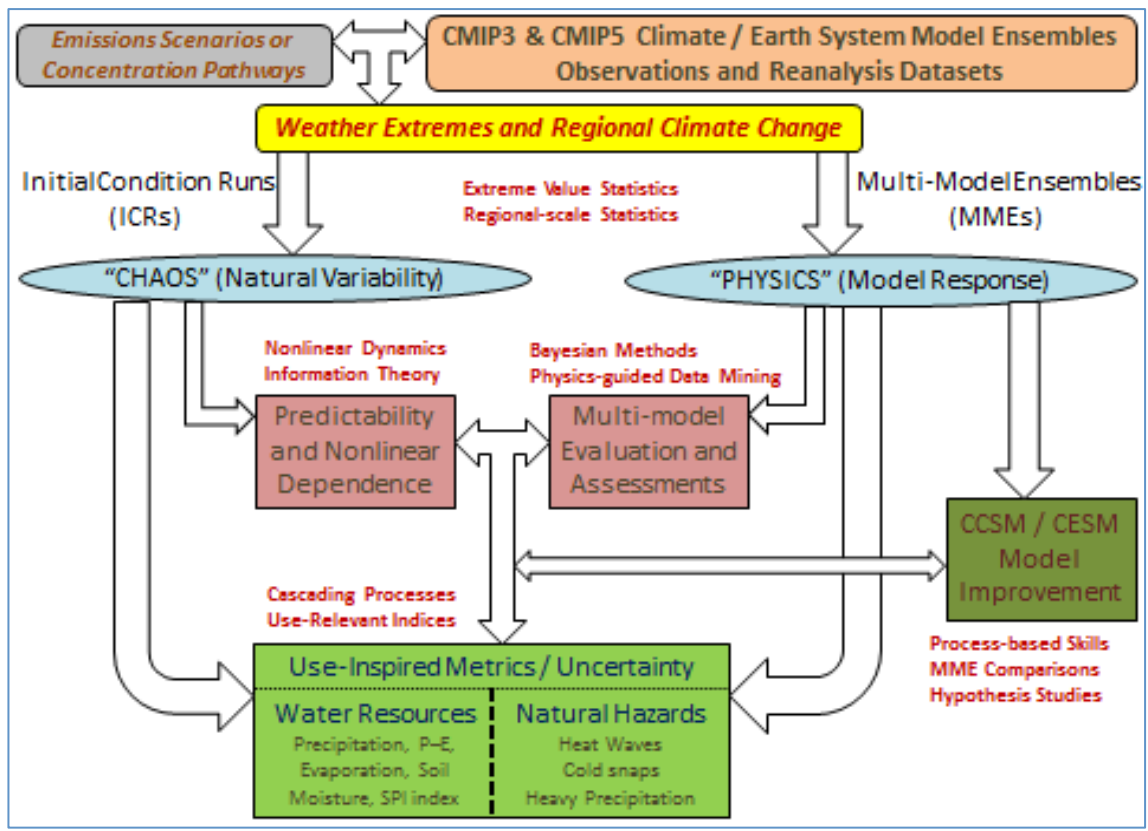

**Figure 6.4** Components, methods, and flowchart of task descriptions, for proposed uncertainty characterization of IESMs. The focus will be on extreme weather and regional climate change, particularly with multi-model and initial condition runs. The stakeholders will be from both climate (IESM) modelers or single-model UQ experts and natural hazards or water resource managers.

### 6.2.3 Combined Impacts of Low Flows and High Stream Temperatures

Power production at risk could be due to the inadequate water supply or high water temperature or both. So far in this proof of concept analysis work we have mostly studied the effect of water stressors separately. In the proposed work, we will perform probabilistic risk assessment by considering the combined effects of water supply and stream temperature focusing on hot spots (regions that will be exposed to both low flows and high temperatures). In addition, we will also look at the seasonal and monthly stream flows and temperatures. The annual estimates of fresh water may hide the information about low stream flows during summer especially in regions where sufficient amount of precipitation may fall during winter but less during summer.

Northeastern University

SDS

### 6.2.4 Understanding Power Plant Attributes

In the absence of detailed information about the specific power plant operation, the present analysis is mostly qualitative and gives a first order indication about which existing power plants and how much electricity production is at risk due to water stress. For a rigorous assessment, we would include additional data such as water withdrawal. The importance of water withdrawals in operational risk assessment of power production is discussed in the NGS Water Intake Project[7] (2005), which indicates withdrawal during low flows is a serious concern. Further, we have considered the availability of fresh water in power plant risk assessment notwithstanding the fact that in coastal areas the use of saline water instead of fresh water expands overall available water supply for power production (Kenny et al., 2009). In this aspect we would need more information about the power plants in detail such as the location of intake pipes, source of water for individual power plants and their consumptive usage[8]. We look forward to collaborate with ARPA-E for future work.

### 6.2.5 Trans-boundary Water Requirement

In the present analysis, we assessed risk as an exposure to water stress due to a combination of various factors such as climate change, population growth, increasing trends in stream temperature specific to each county. However, we ignored the possibility of electricity imports and exports between counties. This simplification results in underestimation of supply-chain risk to surrounding counties. National data show that only ~ 7% counties have power plants with more than 1 MW nameplate capacity, which indicates ~ 93% of the counties have limited endogenous power generation capacity and hence rely on electricity imports from surrounding regions (Cohen and Ramaswami, 2014). Thus trans-boundary water requirements help to elucidate the extent to which a particular county is vulnerable to drought not only in terms of water supply but also energy supply from neighboring counties.

---

[7] Navajo Generating Station extended its cooling water intake pipes in 2005, as water levels in Lake Powell fell due to an extended drought (NGS Water Intake Project, 2005)

[8] Consumptive water usage for power plant refers to the water lost to the environment by evaporation, transpiration, or incorporation into the product (Torcellini et al. 2003).

Northeastern University

SDS

# Appendix

A1. Efficiency of Thermoelectric Power Plants

Table A1. State-of-the-art literatures on assessing power production at risk due to water availability

Table A2. Best practices to estimate water temperature

Table A3. List of major water resources regions in the Contiguous United States

Table A4. Allowable limits of water temperature in various states

Table A5. Details of the stream gauge locations with significant rise (drop) in stream temperature

Table A6. List counties and states that exceeds EPA regulations for stream temperature during projected time period

## A1. Efficiency of Thermoelectric Power Plants

The water requirements for open loop facility is expressed as (van Vliet et al., 2012; Bartos and Chester 2014)

$$q = P.\frac{1-\eta_{total}}{\eta_{elec}}\text{g}\frac{(1-\alpha)}{\rho_w C_p \max\left(\left(\min\left(T_{max}-T_w\right),\Delta T_{max}\right),0\right)} \tag{1}$$

Where, $q$ = required water withdrawal of the power plant (m$^3$/s), $P$ = Installed capacity (kW)

$\eta_{total}$ = net plant efficiency, $\eta_{elec}$ = electrical efficiency, α = share of waste heat not discharged by cooling water, $\rho_w$ = density of liquid water, $C_p$ = heat capacity of water

$T_{max}$ = maximum permissible intake water temperature (°C), $T_w$ = ambient stream temperature (°C), $\Delta T_{max}$ = maximum permissible temperature rise of water (°C)

The maximum usable capacity for the open loop facility is determined using following formula (van Vliet et al., 2012; Bartos and Chester 2014),

Northeastern University

SDS

$$P_{max} = \frac{\min\left((\gamma \cdot Q), q\right) \cdot \rho_w \cdot C_p \cdot \max\left(\min\left((T_{max} - T_w), \Delta T_{max}\right), 0\right)}{\frac{1 - \eta_{total}}{\eta_{elec}} \cdot \lambda \cdot (1 - \alpha)} \tag{2}$$

Where, $\gamma$ = maximum fraction of streamflow available for power generation, $Q$ = natural stream flow, $\lambda$ = correction factor considering changes in efficiency. In Eq. 2, as $T_w$ approaches $T_{max}$, once-through cooling systems must withdraw additional water to maintain same generating capacity. If sufficient additional water is not available then usable plant capacity reduced (van Vliet et al., 2012). Further, when $T_w \geq T_{max}$ plant has to shut down completely (van Vliet et al., 2012).

In recirculating system water is recycled through condenser multiple times by using cooling tower or cooling pond to facilitate heat transfer via evaporative cooling (*consumption*). The water that is not evaporated during cooling process is re-used, means much less water is withdrawn. The water requirements and capacity of recirculating cooling system must be modified to account for the effect of (1) water re-use; and (2) additional climatological and physical constraints (Bartos and Chester 2014). Recirculating cooling system rejects heat primarily through latent evaporative cooling. When air temperature is high, the performance capacity of the plant reduces. The water demand and usable capacity of recirculating cooling plants can be described using following equations (van Vliet et al., 2012; Bartos and Chester 2014),

$$q = P \cdot \frac{1 - \eta_{total}}{\eta_{elec}} \cdot \frac{(1 - \alpha) \cdot (1 - \beta) \cdot \omega \cdot \varepsilon}{\rho_w \cdot C_p \cdot \max\left(\min\left((T_{max} - T_w), \Delta T_{max}\right), 0\right)} \tag{3}$$

$$P_{max} = \frac{\min\left((\gamma \cdot Q), q\right) \cdot \rho_w \cdot C_p \cdot \max\left(\min\left((T_{max} - T_w), \Delta T_{max}\right), 0\right)}{\frac{1 - \eta_{total}}{\eta_{elect}} \cdot \lambda \cdot (1 - \alpha) \cdot (1 - \beta) \cdot \omega \cdot \varepsilon} \tag{4}$$

Where, $\beta$ = fraction of waste heat released into the air, $\omega$ = correction factor to adjust changes in air temperature and humidity and $\varepsilon$ = densification factor considering blowdown (removal) due to increase in salt content in recirculating water. In general as ambient air temperature increases, the heat transfer rate of the plant increases and the power output decreases. Using regression model Colman (2013) showed that with every 1°C rise in air temperature results in 0.01% decrease in plant efficiency, while every 1°C increase in stream temperature leads to 0.02% decrease in plant efficiency, though these vary depending on the generating technology and cooling system type. Recent record shows ~ 53% power plants are recirculating type ~ 43% are once through type[9]. Water withdrawal of once through cooling systems is about 30-50 times higher than the closed circuit cooling systems (Feeley III et al., 2008). Although withdrawal to consumption ratio is significantly lower for recirculating system, total evaporation reduces the thermal efficiency by increasing the salt concentration (Chandel et al., 2011).

---

Northeastern University

**Table A1.** State-of-the-art literature on assessing power production at risk due to water availability

| Citation | Methodology | Case Study | Remarks |
|---|---|---|---|
| Roy et al., (2005) | Available renewable water: $1/n \sum_{i=1}^{n} (P - PET)$, $n$ = 1, ..., $n$ year at a time step of months; where $P$ = precipitation, $PET$ = potential evapotranspiration | **Regions**: Continental U.S. <br><br> **Scenario**: Two scenarios: <br><br> 1. BAU[1]: water use rates remain same as 1995 values over 2000 to 2025. <br><br> 2. Improved Efficiency: doubling power production rate over 1975 – 2000 with no change in freshwater withdrawal. <br><br> **Data:** 1. Precipitation and $PET$ from NOAA[2] <br><br> 2. Water use: USGS data set (Solley et al., 1998) <br><br> 3. Power plant: EIA-906 (EIA[3], 2003) <br><br> **Historical Period**: 1934 – 2002; **Projection Period**: 2025 <br><br> **Assumptions/Caveats**: 1. Irrigation, livestock, mining, industrial and commercial water withdrawals were assumed to be same as 1995 estimate. <br><br> 2. Increase in demand of water is due to population and electricity production. | **Domain**: County <br><br> **Specific Findings:** <br><br> 1. The Southwest and major metropolitan areas in U.S. are likely to have significant new storage requirement. |
| Sun et al. (2008) | Water Supply Stress Index (WaSSI) = $\frac{WD_i}{SW_i + GW_i}$ <br><br> Where, $i$ = individual watershed; $WD_i$ = annual water withdrawal; $SW_i$ = annual water supply for surface water; $GW_i$ = annual supply for groundwater <br><br> • Higher WaSSI indicate greater water stress. WaSSI > 1 indicate demand exceeds supply. <br><br> • Land use changes are simulated through land use change model developed for U.S. South (Hardie et al., 2000) | **Region**: 13 Southeast states: West Virginia, Virginia, Kentucky, Tennessee, North and South Carolina, Arkansas, Oklahoma and Gulf Coast of U.S. <br><br> **Scenario**: <br><br> **Data:** 1. Climate model: HadCM2Sul model, developed by U.K. Hadley Climate Research Center; and CGC1 model, developed by Canadian Climate Center <br><br> 2. Meteorological data: Gridded monthly precipitation and air temperature data are obtain from VMAP[4] (Kittel et al., 1997) <br><br> 3. Water use: USGS data set (Solley et al., 1998) <br><br> 4. Population: historical, U.S. Census Bureau; projected, NPA data services (1999) <br><br> **Baseline period**: 1985 – 1993, **Projection Period**: 2020 | **Domain**: watershed <br><br> **Specific Findings:** <br><br> 1. Population growth result in significant water stress in Piedmont region (Atlantic coastal plain) and Florida. <br><br> 2. Climate change affected regional water supply and demand in Western Texas region. |

| Citation | Methodology | Case Study | Remarks |
|---|---|---|---|
| Harto and Yan (2011) | Loss of power generation is linked to hydrological drought in a river basin.<br><br>Loss of thermoelectric generation =<br><br>At risk thermo ×<br><br>$\left(1 - \left(\frac{drought\ flow}{Min(mean\ flow, water\ demand\ during\ 2010)}\right)\right)$<br><br>Where, At risk thermo = Generation capacity of thermoelectric plants in MW | **Regions**: Eight water resources regions in HUC[5]-2 and HUC-8.<br><br>Regions include, Pacific Northwest, Great Basin, California, Upper Colorado, Lower Colorado, Missouri, Rio Grande and Texas-Gulf<br><br>**Scenario**: Three drought scenarios:<br>1. 10[th] percentile low flow condition in River basins<br>2. 1977 drought year<br>3. Flow condition during 2001<br><br>**Data:** Stream flow data from USGS WaterWatch website (http://waterwatch.usgs.gov/)<br><br>**Historical Period**: 1901-2009<br>**Assumptions/Caveats**: 1. First order worst-case scenario analysis.<br>2. Only average annual losses are considered.<br>3. Does not consider impact of shorter term, extreme events.<br>4. Does not consider cumulative impact of extended severe drought. | **Domain**: basin<br><br>**Specific Findings:**<br><br>1. The Pacific Northwest and Texas basins are found to be vulnerable.<br><br>2. Geospatial correlation show how dependence between power generation and surface water withdrawal for power generation at Pacific Northwest. |
| EPRI (2011) | Water supply sustainability risk index (WSSRI): Set of metrics considered (Roy et al., 2003):<br><br>• Extent of development of available renewable water: greater than 25% of available $P$<br>• Sustainable groundwater use: groundwater withdrawal/total withdrawal greater than 25%<br>• Susceptibility to drought: Summer deficit = Water withdrawal – precipitation; greater than 10 inches during dry months (July, August, September)<br>Water withdrawal for irrigation $\propto P - PET$<br>• Growth in water demand: Increase of total freshwater withdrawal between 2030 and 2005 is more than 20%<br>• Increased need for storage: Summer deficit increases more than 1 inch over 2005 and 2030 | **Regions**: 13 EMM[6] regions<br><br>**Scenario**: A set of scenarios:<br><br>1. BAU: Rate of water use remain at their 2005 level<br><br>2. Increased Efficiency: Thermoelectric water use rate and municipal water use rate is assumed to decrease by 50% and 25% over 25-year period.<br><br>3. Renewable Intensive: Demand in 2030 is met through non-thermoelectric sources with minimal (50 gallon/MW-hr) water consumption.<br><br>4. Partial once through cooling conversion: Conversion of existing once through cooling system to recirculating system.<br><br>5. Fixed withdrawal: withdrawal in 2030 does not exceed current (2005) withdrawal<br><br>**Data:**<br>1. Population: U.S. census Bureau<br>2. Water use: USGS data set 1970-2005 (Kenny et al. 2009).<br>3. Power plant data: EIA – 767 (2009)<br><br>**Historical Period**: 1950 – 2005; **Projection Period**: 2006 - 2030<br><br>**Assumptions/Caveats**: Future water requirement till 2030 is estimated based on linear extrapolation of current trend. | **Domain**: County<br><br>**Specific Findings:**<br><br>1. Significant stresses in Southern/Southwestern, and great plains<br><br>2. ~ 250,000 MW existing power are at risk |

| Citation | Methodology | Case Study | Remarks |
|---|---|---|---|
| Chandel et al. (2011) | Water use factor (WUF) = $\frac{\text{water use by cooling system}}{\text{generator capacity}}$<br><br>1. WUF = $f(\boldsymbol{\theta})$, where $f$ = Least square regression<br><br>$\boldsymbol{\theta}$ = {type of cooling systems; type of fuel, thermal efficiency, operational conditions, and type of water source}<br><br>2. Separate regression models are developed for each cooling type. | **Regions**: 13 NERC[7] regions<br>**Scenario**: BAU<br>**Data:** Water use and power plant information from EIA-767, EIA-860 and NETL[8] Coal Plant database<br>**Historical Period:** 1996 – 2005; **Projection period**: 2010 – 2030<br>**Assumptions/Caveats**: Evaporative losses are not considered. | **Domain**: Individual power plant<br>**Specific Findings:**<br>1. Climate policies and carbon price may reduce electricity generation and increase water consumption if existing coal plants are not installed with $CO_2$ capture facilities. |
| Roy et al., (2012) | Water supply sustainability risk index | **Region**: Continental U.S.<br>**Scenario**: A1B<br>**Data:** 1. Climate: 16 Global Climate Models from CMIP3 archive<br>2. Population: U.S. census Bureau<br>3. Water use: Kenny et al. (2009).<br>4. Power plant data: EIA report (2009).<br>**Historical Period:** 1934 - 2005<br>**Projection period:** 2050 (averaged over 2040 – 2059)<br>**Assumptions/Caveats**: (i) Water use rate is assumed to increase due to domestic supply and thermoelectric cooling.<br>(ii) Water use in other sector is assumed to be same as in 2005.<br>(iii) PET is estimated using Hamon's equation, which is more sensitive to temperature | **Domain**: County<br>**Specific Findings:**<br>1. Water supplies in 70% of counties are at risk due to climate change.<br>2. ~ 1/3$^{rd}$ of counties are at extreme risk; regions include California, Nevada, Arizona, Texas and parts of Florida. |
| Scanlon et al. (2013) | Water Intensity = $\frac{\text{water use by cooling system}}{\text{Net generation}}$<br><br>Where, Water use = includes consumption or withdrawal<br><br>Net generation = gross generation – electrical energy consumed at the generating stations for station service or auxiliaries<br><br>• Power plant discharge temperature | **Region**: Texas<br>**Scenario**: BAU; Water intensity during 2011 (drought year) is compared with 2010 (non-drought year).<br>**Data:** Water use data from TCEQ[9], TWDB[10]<br>**Baseline period:** 2010; **Projection period**: 2030 | **Domain**: Individual power plant<br>**Specific Findings:**<br>1. Western Texas is well adapted to drought by using alternate water sources (groundwater and municipal waste water) as compared to relatively humid eastern Texas. |

| Citation | Methodology | Case Study | Remarks |
|---|---|---|---|
| Averyt et al., (2013) | Water Supply Stress Index (WaSSI)<br><br>Surface runoff is simulated by running a distributed hydrological model. | **Region:** Major hydrologic unit<br><br>**Data:** 1. Water use by thermoelectric sector from EIA-860 and EIA-923<br><br>2. surface and Groundwater withdrawal data is based on county level USGS data for 2005 (Kenny et al. 2009)<br><br>3. Projected runoff data from Milly et al. (2005).<br><br>**Scenario:** A1B<br><br>**Historical Period:** 1999 – 2007; **Projection period:** 2041 – 2060<br><br>**Assumptions/Caveats:** 1. Irrigation return flows are not considered.<br><br>2. WaSSI does not consider the volume of groundwater remaining and assume unlimited groundwater supplies. | **Domain:** watershed scale<br><br>**Specific Findings:**<br><br>1. Western U.S. is sensitive to low flow events and projected long-term shifts due to climate change.<br><br>2. A few parts of Southeast are also vulnerable.<br><br>3. Demand of freshwater exceeds natural supplies over 9% of the total watershed examined. |
| Blanc et al., (2014) | Water Stress Index (WSI) Smakhtin et al. (2004)<br><br>$$= \frac{WD}{MAR - EWR}$$<br><br>$WD$ = Mean Annual Withdrawal<br><br>$MAR$ = Mean Annual Runoff<br><br>$EWR$ = Environmental Water Requirement<br><br>Heavily exploited: $0.6 \leq WSI \leq 1$; Overexploited: $WSI > 1$<br><br>• **Simulation Framework:** MIT Integrated Global System Model (IGSM) for U.S.:<br>1. Runoff: Community Land Model (CLM) ver. 3.5<br><br>2. Crop Water Requirement: CliCrop<br><br>3. Greenhouse gas (GHG): Emissions Prediction and Policy Analysis (EPPA) model<br><br>4. Region Specific Economic Activity: U.S. Regional Economic and Environmental Policy (USREP) model<br><br>5. Energy: Regional Energy Deployment System (ReEDS)<br><br>6. Water balance and Water Stress: Water System Management (WSM) module<br><br>• Inter-basin water transfer is considered<br>• Climate variables are downscaled at power plant scale | **Region:** Conterminous U.S.<br><br>**Scenario:** Two GHG scenarios are considered:<br><br>• Unconstrained Emissions (UCE): No change in GHG emissions<br>• Level 1 Stabilization (L1S): $CO_2$ concentration is equivalent to 450 ppm.<br><br>**Historical Period:** 2005 – 2009; **Projection period:** 2041 – 2050<br><br>**Data:**<br><br>1. Climate models: Two climate models; GFDL2.1, CCSM3[11] from CMIP3 archive<br><br>2. Groundwater use: Kenny et al. (2009)<br><br>3. Streamflow data: 99 Assessment Sub-regions delineated by U.S. Water Resources council<br><br>4. population: U.S. Census Bureau<br><br>**Assumptions/Caveats:** 1. Irrigated areas remain unchanged in future.<br><br>2. Water withdrawals are estimated annually. Monthly values are assumed as evenly spread across year.<br><br>3. Water restriction posed by a region during dry period is not considered. | **Domain:** basin<br><br>**Specific Findings:**<br><br>1. Population and economic growth are the major factor in increasing water stress in U.S. through mid-century.<br><br>2. Climate change increases the water stress in Southwest. |

*Note:* [1]BAU: Business as usual; [2]NOAA: National Oceanic and Atmospheric Administration; [3]EIA: Energy Information Administration; [4]VMAP: Vegetation/Ecosystem Modeling and Analysis Project; [5]HUC: Hydrologic Unit Code; [6]EMM: Electric Market Module; [7]NERC: North American Electric Reliability Corporation; [8]NETL: National Energy Technology Laboratory; [9]TCEQ: Texas Commission on Environmental Quality; [10]TWDB: Texas Water Development Board; [11]CCSM3: Community Climate System Model ver. 3

**Table A2.** Best practices to estimate water temperature

| Approach | Citation | Temporal Scale | Case Study/Application | Specific Findings |
|---|---|---|---|---|
| *Functional Approximation* | Pilgrim et al. (1998) | Daily, weekly, monthly and annual | **Study Area**: Minnesota<br><br>**Model**: Linear regression<br><br>**Predictor(s):** Air temperature<br><br>**Data:**<br><br>1. Stream temperature: USGS stream database<br>2. Air temperature: Midwest Climate Center of the Illinois State Water Survey, Champaign, Illinois<br><br>**Number of Stations:** Stream gauges 39; Weather station 39<br><br>**Time Period**: *Historical* 1956 – 1991 varies from stream to stream; *Projected* $2{\times}CO_2$ climate condition from GISS and GFDL scenarios<br><br>**Assumption/Caveats**:<br><br>1. No time lags were considered while developing regression relationship.<br><br>2. Linear regression between stream and air temperature is only accurate at moderate air temperature (e.g., 0 to 20°C).<br><br>3. Periods of ice cover (November to March) were excluded since regression equations do not cover air temperature below 0°C | On average stream temperature were projected to rise by 4.1°C in warm season (April – October). |
| | Mohseni et al. (1999) | Weekly | **Study Area**: Conterminous U.S.<br><br>**Model**:<br>1. Four parameter non-linear Logistic regression model.<br>2. Separate models for warm and cool season.<br>3. One parameter (the upper bound) is obtained via extreme value analysis and remaining three by least square regression<br><br>**Predictor(s):** Air temperature<br><br>**Data:**<br><br>1. Stream temperature: U.S. EPA[1], Mid-Continent Division, Duluth, Minnesota<br>2. Air temperature: NOAA, NREL[2]<br><br>**Number of Stations:** Stream gauges 803; Weather stations 166<br><br>**Time Period**: Historical 1961 – 1979; Projected $2{\times}CO_2$ climate condition from CCC-GCM[3]<br><br>**Assumption/Caveats**: The stream temperature and air temperature assumed to follow an S-shaped relation | 1. 764 stream gages projected an increase in mean annual stream temperature by 2 - 5°C, least near the West Coast and most in the Missouri and Ohio River basin.<br><br>2. Only 39 stream gauges exhibit insignificant changes in projected climate state.<br><br>3. 1 - 3°C increase in maximum and minimum stream temperature is noted in Central U.S.<br><br>4. Most streams exhibit maximum changes in weekly stream temperature during spring.<br><br>5. Minimum changes in stream temperature is observed during winter (December – January) and summer (July – August) throughout the U.S. in projected climate condition. |

| Approach | Citation | Temporal Scale | Case Study/Application | Specific Findings |
|---|---|---|---|---|
| *Functional approximation* | Sahoo et al. (2009) | Daily | **Study Area**: Lake Tahoe, CA-NV border<br><br>**Model**: MRA[4], ANN[5], and CNDA[6]<br><br>**Predictor(s):** Air temperature and solar radiation at different time lags<br><br>**Data:** USGS database<br><br>**Number of Stations:** 4 Streams that inflows to Lake Tahoe<br><br>**Time Period**: Training 1/1/1999 – 12/31/2001<br>Testing 2002 | 1. Prediction performance of ANN-based model is found to be highest.<br><br>2. Sensitivity analyses indicate air temperature as the most important variable in stream temperature prediction.<br><br>3. Inclusion of short-wave radiation improve the prediction performance, however shortwave radiation alone could not predict stream temperature with reasonable accuracy. |
| | Jeong et al. (2013) | Daily | **Study Area**: Ouelle River basin, Québec, Canada<br><br>**Model**: Mixed model<br><br>Stream temperature = $f$ (seasonal and residual component)<br><br>Where $f$ denotes summation; Seasonal component = sine function; residual component = ANN based model<br><br>**Predictor(s):** Air temperature<br><br>**Data:**<br><br>1. Stream temperature: Observed stream temperature based on temperature loggers<br>2. Air temperature: Station observed data from weather station La Pocatiére, Canada<br>3. Climate Model Output: 5 regional climate model simulations from NARCCAP[7] in SRES emission scenarios (B1, A1B and A2)<br><br>**Number of Stations:** 18 observation sites (11 main stream and 7 tributary)<br><br>**Time Period**: Training 1970 – 1999; Projection 2046 – 2065<br><br>**Assumption/Caveats**:<br>Future changes in autocorrelation structure of water temperature are not tested. | MIROC (A1B scenario) projected largest increase in the summer mean water temperature |
| *Hydrological model* | SHADE-Hydrologic Simulation Program – Fortran (HSPF) (Chen et al., 1998a, 1998b) | Hourly | **Application:** Upper Grande Ronde watershed in northeast Oregon<br><br>**Model**:<br>• Analyzes stream temperature in a watershed by solving an unsteady heat conduction equation.<br>• Dynamically adjust solar radiations taking into account sun position, stream location, orientation and riparian shading characteristics adjacent to streams. | |

| Approach | Citation | Temporal Scale | Case Study/Application | Specific Findings |
|---|---|---|---|---|
| *Hydrological Model* | Stream Network Temperature Model (SNTEMP) (Bartholow, 2000) | Daily | **Model**: 1-D heat transport model; Simulates mean and maximum water temp as a function of stream distance and environmental heat flux. The tool is available at USGS website (https://www.fort.usgs.gov/products/10016)<br><br>Specific Features:<br>• Applicable for stream network of any size and order<br>• Corrects air temperature, relative humidity, and atmospheric pressure as a function of elevations<br>• Fills missing observations in stream temperature data<br>• Uses time steps ranging from 1 month to 1 day<br><br>**Assumption/Caveats**:<br>• Inability to deal with rapidly fluctuating flows<br>• Uses an empirical approach to predict maximum daily stream temperature. For example, maximum afternoon air temperature in the model is a function of radiation, humidity, sunshine hours and a set of empirically derived coefficients (Bartholow, 2000) | |
| | Yearsley (2009) | Daily | **Application:** Pacific Northwest<br><br>**Model**: Semi-Lagrangian numerical method that solves 1-*d* heat transport equation<br><br>**Data:** Stream temperature and flow data are archived at USGS and DART data site (http://www.cbr.washington.edu/dart/)<br><br>**Number of Stations:** 6 dam locations at Clearwater, Snake and Columbia River basins<br><br>**Time Period**: Baseline 1995 – 2000; Projection 2020, 2040<br><br>**Assumption/Caveats**:<br>• Applicable for small and medium sized catchments<br>• Simulates at smaller temporal scale (*i.e.*, hourly, sub-daily and daily)<br>• Neglects topographic and riparian shading<br>• Neglects streambed and hyporheic[*] heat exchange | |
| | Hydrologic Engineering Center – River Analysis System (HEC-RAS) ver. 4.0 (Jensen and Lowney, 2004) | Daily | **Application:** Sacramento River, North Carolina<br><br>**Model:** Numerically solves 1-*d* energy budget approach.<br><br>Model input includes hydrodynamic information, system geometry, temperature at hydrodynamic boundaries and meteorological data.<br><br>**Assumption/Caveats**:<br>• Neglects topographic and riparian shading<br>• Neglects streambed and hyporheic heat exchange | |

[*]Hyporheic exchange refers to the phenomena when surface water enters the shallow subsurface, e.g., channel bed, banks or morphological features and then reemerges back into the main channel  (Burkholder et al., 2008)

| Approach | Citation | Temporal Scale | Case Study/Application | Specific Findings |
|---|---|---|---|---|
| *Hydrological Model* | van Vliet *et al.*, (2013) | Daily | **Study Area**: Global<br><br>**Model**: VIC-RBM modeling framework: VIC captures the hydraulic characteristics of river, while RBM uses output from VIC and solves 1-$d$ heat transport equation (Yearsley, 2009).<br><br>Spatial Resolution: 0.5°<br>**Data**: 1. Stream Temperature: GEMS[8] database.<br>Climate Models: Output of 3 bias-corrected GCM output (SRES A2 and B1 emission scenarios): CNRM-CM3, IPSL-CM4, and ECHAM5 in CMIP3 archive<br>2. Thermoelectric Water Use: Global gridded thermoelectric water use database (Vassolo and Döll, 2005)<br>**Number of Stations**: 333 stations globally<br>**Time Period**: Baseline 1971 – 2000; Projection 2071 – 2100<br>**Assumption/Caveats:**<br>• Does not consider effect of topographic and riparian shading<br>• Does not consider streambed and hyporheic heat exchange. | 1. Global mean and extreme (95th percentile) stream temperatures are projected to increase on average ~0.8 – 1.6°C relative to baseline period.<br><br>2. The largest stream temperature increases are projected for the U.S., Europe, Eastern China, and parts of Southern Africa and Australia.<br><br>3. Large increase in stream temperature accompanied by low flows is projected for Southeastern U.S., Europe, eastern China, Southern Africa and Southern Australia. |
| *Trend Analysis* | Kaushal et al. (2010) | Annual | **Study Area**: Conterminous U.S.<br>**Model**: linear regression and non-parametric Mann-Kendall trend tests with Sen's slope estimate<br>**Data:**<br>Stream temperature: USGS database; Air temperature: USHCN[9]<br>**Number of Stations:** Stream gauges 40<br>**Time Period**: 24 to 100 years record (1908 – 2007) depending on period of data availability | 1. 20 major streams showed statistically significant long-term warming.<br><br>2. Long-term increase in stream temperature is correlated with increases in air temperature. |
|  | Luce et al. (2014) | Weekly | **Study Area**: Pacific Northwest – Oregon and Washington.<br><br>**Model**: Principle component analysis (PCA) and MRA at each stream temperature station:<br>• PCA to reconstruct air temperature and streamflow time series<br>• Reconstructed time series is regressed against stream temperature<br>**Data**:<br>stream temperature: U.S. forest service database; Air temperature: USHCN<br>**Number of Stations:** Stream gauges 246; Air temperature stations, 25<br>**Time Period**: Average 10 years (1988 – 2010) of record depending on the data availability | 1. At inter-annual time scales, colder streams are less sensitive to air temperature fluctuations than warmer streams. This pattern holds good for direct warming through radiative heat transfer.<br><br>2. Stream temperature is also sensitive to secondary influential factors, *e.g.*, riparian disturbance (fire/debris flows), earlier snowmelt (decreased summer flows) and reduced groundwater recharge. |

**Note**: [1]EPA: Environmental Protection Agency; [2]NREL: National Renewable Energy Laboratory; [3]CCC-GCM: Canadian Center of Climate Modeling – General Circulation Model; [4]MRA: Multiple Regression Analysis; [5]ANN: Artificial Neural Network; [6]CNDA: Chaotic Non-linear Dynamic Algorithm; [7]NARCCAP: North American Regional Climate Change Assessment Program; [8]GEMS: Global Environment Monitoring System; [9]USHCN: U.S. Historical Climatology Network

**Table A3.** List of major water resources regions in Conterminous U.S.

| HUC # | Regions analyzed |
|-------|------------------|
| 01 | New England |
| 02 | Mid Atlantic |
| 03 | South Atlantic-Gulf |
| 04 | Great lakes |
| 05 | Ohio |
| 06 | Tennessee |
| 07 | Upper Mississippi |
| 08 | Lower Mississippi |
| 09 | Souris – Red – Rainy |
| 10 | Missouri |
| 11 | Arkansas – White - Red |
| 12 | Texas - Gulf |
| 13 | Rio Grande |
| 14 | Upper Colorado |
| 15 | Lower Colorado |
| 16 | Great Basin |
| 17 | Pacific Northwest |
| 18 | California |

*Note:* HUC denote hydrologic unit code

Northeastern University

SDS

**Table A4.** Allowable limits of water temperature in various states

| State | Temperature Threshold (°C) |
|---|---|
| Indiana | 35 |
| Kentucky | 31.7 |
| Louisiana | 34.4 |
| North Carolina | 34.8 |
| Pennsylvania | 30.5 |
| Virginia | 33.7 |
| Wisconsin | 31.7 |

*Note:* The allowable temperature threshold for all other states is 32.2°C. (*Source*: Madden et al., 2013)

Northeastern University

SDS

**Table A5.** Details of the stream gauge locations with significant rise (drop) in stream temperature

| Trend | State | County | Start Year | End Year | Number of Years |
|---|---|---|---|---|---|
| Upward | Massachusetts | Middlesex | 2004 | 2012 | 9 |
| | Pennsylvania | Carbon | 1981 | 2002 | 22 |
| | New Jersey | Mercer | 1996 | 2008 | 13 |
| | Pennsylvania | Chester | 1997 | 2010 | 14 |
| | Pennsylvania | Schuylkill | 2000 | 2007 | 8 |
| | Pennsylvania | Schuylkill | 1997 | 2007 | 11 |
| | North Carolina | Caswell | 1998 | 2012 | 15 |
| | North Carolina | Martin | 1998 | 2012 | 15 |
| | North Carolina | Beaufort | 2000 | 2009 | 10 |
| | South Carolina | Saluda | 1997 | 2012 | 16 |
| | South Carolina | Berkeley | 1999 | 2012 | 14 |
| | Florida | Putnam | 1998 | 2007 | 10 |
| | Florida | Orange | 1983 | 2001 | 19 |
| | Florida | Osceola | 1978 | 2008 | 31 |
| | Georgia | Gwinnett | 2003 | 2012 | 10 |
| | Kentucky | Edmonson | 2000 | 2012 | 13 |
| | Tennessee | Clay | 1992 | 2012 | 21 |
| | North Carolina | Bedford | 1992 | 2012 | 21 |
| | New York | Monroe | 1995 | 2011 | 17 |
| | North Dakota | Traill | 1999 | 2012 | 14 |
| | Wisconsin | Dane | 1997 | 2006 | 10 |
| | Montana | Madison | 1996 | 2012 | 17 |
| | Montana | Broadwater | 1978 | 2012 | 35 |
| | Colorado | Pueblo | 1988 | 2012 | 25 |
| | Texas | Armstrong | 1969 | 1991 | 23 |
| | Texas | Harris | 1988 | 2000 | 13 |
| | New Mexico | Sandoval | 1972 | 1982 | 11 |
| | Arizona | AZ | 1994 | 2012 | 19 |
| | Arizona | Coconino | 1993 | 2000 | 8 |
| | Montana | Missoula | 1995 | 2012 | 18 |
| | Pennsylvania | Schuylkill | 1997 | 2007 | 11 |
| | North Carolina | Beaufort | 2001 | 2009 | 9 |
| | North Carolina | Graham | 2000 | 2012 | 13 |
| | Washington | Skamania | 1998 | 2012 | 15 |
| Downward | Wyoming | Park | 2004 | 2012 | 9 |
| | Wyoming | Park | 2003 | 2010 | 8 |
| | Oregon | Douglas | 2004 | 2012 | 9 |

**Table A6.** List counties and states that exceeds EPA regulations for stream temperature during projected time period

| Sl. no | 2030 | | 2040 | |
|---|---|---|---|---|
| | County | State | County | State |
| 1 | Beaufort | South Carolina | Berkeley | South Carolina |
| 2 | Plaquemines Parish | Louisiana | Beaufort | South Carolina |
| 3 | Houston | Texas | Adair | Kentucky |
| 4 | Chambers | Texas | Plaquemines Parish | Louisiana |
| 5 | | | Skamania | Washington |
| 6 | | | Houston | Texas |
| 7 | | | Chambers | Texas |

Northeastern University
SDS

Northeastern University

Northeastern University

Northeastern University

Northeastern University

SDS

Northeastern University

SDS

Northeastern University

SDS
SUSTAINABILITY & DATA SCIENCES LAB

Northeastern University

SDS